%% file: WNflows.tex
\documentclass[reprint,aps,twocolumn,prl,preprintnumbers, nofootinbib,10pt]{revtex4-2}
\input{preambleLetter}
\usepackage{xpatch}
\pdfoutput=1
\usepackage{color}
\usepackage{comment}
\usepackage{enumitem}
\usepackage{mathtools}
\usepackage[dvipsnames]{xcolor}

\begin{document}
\setcounter{equation}{0}

\title{\boldmath RG flows of minimal $\mathcal W$-algebra CFTs via non-invertible symmetries}

\author{Federico Ambrosino}
\email{federicoambrosino25@gmail.com}      
\affiliation{Perimeter Institute for Theoretical Physics, Waterloo, Ontario N2L 2Y5, Canada }                                                              
\author{Tom\'a\v{s} Proch\'azka }
\email{prochazkat@fzu.cz}
\affiliation{Institute of Physics AS CR, Na Slovance 2, Prague 8, Czech Republic}

\date{\today}
                
\begin{abstract}
\noindent In this letter  we study renormalization group (RG) flows between 2d conformal field theories enjoying extended higher-spin $\mathcal{W}$-symmetry. We propose a new class of RG flows between the diagonal minimal models of $\mathcal{W}_N$-algebra that take the form $\mathcal{W}_N(p,q)\to\mathcal{W}_N(p,kp-q)$. These are obtained by matching the anomalies of the non-invertible symmetry ${\mathrm{Rep}}[SU(N)_{p-N}]$ (and its discrete quotients) that is preserved by special relevant primary fields. 
This large non-invertible symmetry includes the familiar $\mathbb{Z}_N$ symmetry of the minimal models. 
Our new flows furnish a significant generalization of the ones recently found in the case of Virasoro algebra, and include all previously known RG flows of $\mathcal{W}_N$. They have the remarkable property of being uniform in the rank $N$ of the $\mathcal{W}$-algebra.
\end{abstract}

\preprint{}

\maketitle

\paragraph{\textbf{Introduction.}}
The behavior of a quantum system  
depend dramatically on the energy scale at which they are probed. In Quantum Field Theory, this phenomenon is understood as a renormalization group (RG) flow between the UV (high energy, microscopic)  and IR (low energy, long distance) fixed points. 
Given a microscopic description for the UV fixed point of a physical system in generic dimensions, it is a remarkably hard task to understand its low energy phases. The prototypical situation is a UV description as conformal field theory (CFT).  A non trivial RG flow is triggered by turning on a deformation by one of the relevant  ($\Delta \lesssim d$ ) primary fields. At long distances, if the excitations of the system are gapped, the IR is a topological quantum field theory (TQFT), and if gapless, it is another CFT.  Identifying the specific nature of either of those is often hard. 
The situation is better in $d=1+1$ dimension. There, the finite-dimensional conformal symmetry is enhanced to an infinite dimensional local version: the Virasoro symmetry\cite{Belavin:1984vu}. Thanks to the Virasoro algebra we know many examples of completely solvable 2d CFTs. Yet, even for these solvable models, understanding their RG flows remains elusive. Symmetries are of  paramount help as they provide RG-invariant quantities that have to match along the flow. 
The generalized\cite{Gaiotto:2014kfa}  (0-form) symmetries of a 2 dimensionalCFTs(see \cite{Schafer-Nameki:2023jdn,Shao:2023gho} for recent reviews on generalized symmetries),  are generated by its topological line operators\cite{Oshikawa:1996dj,Petkova:2000ip,Fuchs:2002cm,Fuchs:2003id,Fuchs:2004dz,Frohlich:2004ef,Frohlich:2009gb} . Those may or may not have a group-like fusion, and in general they form a  \textit{non-invertible} fusion category . Topological defects have been proven to  furnish invaluable tools in constraining RG flows and phases of 2 dimensional theories\cite{Fredenhagen:2009tn,Gaiotto:2012np,Chang:2018iay,Komargodski:2020mxz,Gaiotto:2020fdr,Bhardwaj:2023idu,Cordova:2024vsq,Nakayama:2024msv,Ambrosino:2025yug,Antunes:2025huk,Copetti:2024rqj,Copetti:2024dcz,Ambrosino:2025myh,Ambrosino:2025pjj}
In this paper, we will focus on  2d CFTs of a very special kind: rational CFTs (RCFTs). Those are the simplest in the landscape of 2d CFTs and are characterized by having only finitely many primary fields and rational central charge. 
The set of simple topological line operators commuting with full Virasoro algebra is
in one-to-one correspondence with the set of primaries. Hence, this case represents one of the ideal instances where we do know the entire set of (0-form) symmetries of a given theory. The most familiar examples of 
RCFTs are the Virasoro minimal models $\cW_2(p,q)$ with central charge $c = 1- 6(p-q)^2/(pq)$. These models have received renewed attention\cite{Lencses:2023evr,Benedetti:2024utz, Delouche:2024tjf,Katsevich:2024sov, Katsevich:2024jgq, Antunes:2022vtb,Antunes:2024mfb,Tanaka:2025qou,Gaberdiel:2026sfg,Benedetti:2026tpa,Klebanov:2022syt,Scheinpflug:2023osi,Scheinpflug:2025sqn,Fukusumi:2025fir,Benedetti:2026tpa}.
RG glows triggered by special  primary fields of adjoint type, preserve a large non-invertible fusion category $\Rep[SU(2)_{p-2}]$ \cite{Nakayama:2024msv}. Leveraging on the anomaly  matching of this symmetry, the following class of RG flows has been  proposed \cite{Nakayama:2024msv,Ambrosino:2025yug}:
\be
\cW_2(p,kq+i) \xrightarrow{\phi_{[1,2k+1]}} \cW_2(p,kq-i),
\ee
unifying all the previously known RG flows\cite{Martins:1992ht,Martins:1992yk, Zamolodchikov:1987ti,Fioravanti:1996rz,Feverati:1999sr,Fendley:1993wq,Fendley:1993xa,Dorey:2000zb}. 
Virasoro minimal models are members of a much larger class of RCFTs. Indeed, while Virasoro algebra is generated by the Fourier modes of the spin $s=2$ stress-energy tensor, it is very natural to consider algebras generated by currents of increasingly higher integer spins $s=2,3,\cdots, N$\cite{Zamolodchikov:1985wn,Lukyanov:1990tf,Gaberdiel:2012uj,Watts:1992he}. Those are known as $\cW_N$ algebras and also admit minimal minimal rational truncations $\cW_N(p,q)$. Despite being a natural and  richer generalization of the Virasoro models, only sporadic RG flows have been discussed so far in the literature\cite{Poghosyan:2023brb, Lukyanov:1990tf, Vaisburd:1994hg,Poghosyan:2022ecv,Poghosyan:2022mfw}. 

The main result of this letter is the proposal of the following infinitely many new RG flows between the diagonal $\cW_N$ minimal models:
\be\begin{split}\label{WNflows}\cW_{2N}(p,kp+i)&\to \cW_{2N}(p,kp-i)\\
\cW_{2N+1}(p,\tfrac{k+1}{2}p+i) &\to \cW_{2N+1}(p,\tfrac{k+1}{2}p-i) 
\end{split}\ee
triggered by adjoint-type deformations. This is a complete set of $\Rep[SU(N)_{p-N}]$ symmetry-preserving RG flows, extending the $\bZ_N$ symmetry of the $\cW_N$ models. For $\cW_{2N}$ \eqref{WNflows} can be enlarged to $k \in \bZ/2$: in this case, only the smaller $\Rep[(SU(2N)/\bZ_{N})_{p-2N}]$ symmetry is preserved.

Quite remarkably, these flows admit a uniform description independently on the rank $N$ of the $\cW_N$-algebra
\be 
\cW_N(p, q) \to \cW_N(p, kp - q)
\ee
that is unexpected from the first ground. Furthermore, the embedding of these flows into $\cW_\infty$ algebra, suggests that these flows are part of a much more general structure of flows between truncations of $\cW_\infty$. 

The only flows\cite{ Lukyanov:1990tf, Vaisburd:1994hg} with $N>2$ already known in the literature correspond to the case $k=1, i=1$ of flows between the unitary minimal models, and were only found  thanks to integrability. More details on this can be fond in Appendix D.  Our proposal encompass and generalize the known flows, and  does not rely on  integrable structures. This may serve as an illustration of the power of non-invertible symmetries. 

The letter is organized as follows: after  outlining the general strategy, we analyze in detail the case $\cW_3$ and $\cW_4$, before presenting the general argument. Before offering an outlook of possibile future directions, we also propose RG flows between minimal models of distinct $\cW_N$ algebras. We provide more details, general proofs and explicit examples for $\cW_3(p\leq 8),\cW_4(p\leq 8)$ in the Suppemental Materials.
\paragraph{\textbf{General strategy.}}
Symmetries of 2 dimensional quantum field theories are generated by topological line operators. 
In rational CFTs the set of topological defects commuting with the full chiral algebra is the one of Verlinde line operators. These are in 1-to-1 correspondence with the (finite number of) primaries of the RCFT, and their fusion rules are exactly the ones of the corresponding primaries. We indicate the lines by $\cL_\sigma$, using the same label as for the corresponding primary field  $\sigma$. Verlinde line acts on primaries as
\begin{align}\label{eq:ward}
    \cL_{\sigma} \ket{\phi_{\rho}} &= 
    \begin{tikzpicture}[baseline={(0,-0.5ex)}]
        \draw[ thick,red] (0,0) circle [radius=0.6cm];
        \draw[->,red,thick] (0.6,0) arc [start angle=0, end angle=120, radius=0.6cm];
        \fill (0,0) circle [radius=1pt];
            \node at (0.2, -0.2) {$\phi_{\rho}$};
         \node at (-0.9, -0) {$ \color{red} \cL_{\sigma}$};
    \end{tikzpicture} = \frac{S_{\sigma\rho}}{S_{0\rho}} \ket{\phi_\rho} \comma
\end{align} 
where $S_{ab}$ is the modular S-matrix. Equation \eqref{eq:ward}
may be thought of as the Ward identity associated to the (non-invertible) global symmetry generated by $\cL_\sigma$. Consider an RG flow triggered by a relevant deformation of ${\rm CFT}_{\rm UV}$: 
$$ 
{\rm CFT}_{\rm UV} + g_\varphi\int\varphi\comma \qquad h_\varphi < 1$$
If the operator identity
$\label{opid} [\mathcal{L}, \varphi] \ket{\Phi} = 0 
$, diagrammatically
 \begin{align}
    \begin{tikzpicture}[baseline={(0,-0.5ex)}]
        \draw[ thick,red] (0,0) circle [radius=0.6cm];
        \draw[->,red,thick] (0.6,0) arc [start angle=0, end angle=120, radius=0.6cm];
        \fill (0,0) circle [radius=1pt];
            \node at (0.2, -0.2) {$\phi_{\rho}$};
         \node at (-0.3, 0.2) {$ \color{red} \cL_{\sigma}$};
         \draw[ ultra thick,black] (0,0) circle [radius=1cm];
         \node at (1.4, 0.0) {$ \ket{\Phi}$};
    \end{tikzpicture} =
\begin{tikzpicture}[baseline={(0,-0.5ex)}]
        \draw[ thick,red] (-0.2,0) circle [radius=0.6cm];
        \draw[->,red,thick] (0.4,0) arc [start angle=0, end angle=120, radius=0.6cm];
        \fill (0.75,0) circle [radius=1pt];
            \node at (0.65, -0.2) {$\phi_{\rho}$};
         \node at (-0.3, 0.2) {$ \color{red} \cL_{\sigma}$};
         \draw[ ultra thick,black] (0,0) circle [radius=1.cm];
         \node at (1.4, -0.0) {$ \ket{\Phi}$};
    \end{tikzpicture} 
\end{align}
holds inside any correlation function, then the symmetry generated by the line operator $\cL$ is unbroken along the entire flow.
The simpler condition  turns out to be  sufficient\cite{Gaberdiel:2026sfg,Nakayama:2024msv,Ambrosino:2025yug}: 
\be\label{simplecomm}
\mathcal{L} \ket{\varphi} = d_\mathcal{L} \ket{\varphi}\comma \qquad (\ket{\Phi
}= \ket{0})
\ee
Here $d_{\cL} =\bra{0} \cL\ket{0}$ is the quantum dimension of $\cL$. Therefore, if a deformation $\varphi$ preserves a non-trivial set of non-invertible symmetries of the UV theory, the deep IR phase satisfies non-trivial constraints due the existence of a (possibly spontaneously broken) global symmetry. 
In this paper we consider RG flows where the IR phase is gapless,
$$ 
\rm CFT_{UV}   \xrightarrow{+ g_\varphi \int\varphi} CFT_{IR}.
$$In this case, since the topological line remains topological along the RG flow, it must also flow to a topological defect line in the IR CFT.  In particular, the quantum dimension $d_\cL$ is a RG invariant quantity and must be matched in the IR.

Among the other RG-invariant quantities associated to these flows are the 't Hooft anomalies associated with the preserved symmetries. For our purposes, we are concerned with $\bZ_N$ symmetries. In this case, \cite{Lin:2021udi} the 't Hooft anomaly is classified by the group cohomology $H^3(\bZ_N, U(1)) \simeq \bZ_N$, and  is encoded in the spin content of the defect Hilbert space $\mathcal{H}_\cL$ of a symmetry line,
$$ Z_\cL = \sum_{\rho,\sigma} N^{\cL}_{\rho \sigma} \chi_\rho (q) \, \overline\chi_\sigma(\bar q) = {\rm  Tr}_{\mathcal{H}_\cL} q^{h- \frac{c}{24} }\bar{q}^{\bar{h}- \frac{c}{24}}.$$
The spins $s = h -\bar{h}$ of the states in the defect Hilbert space $\cH_{\cL}$  for a $\bZ_N$ symmetry with anomaly class $k \in \bZ_N$ satisfy\cite{Chang:2018iay,Lin:2021udi}
\be 
s = h- \bar{h} \in \frac{k}{N^2} + \frac{\bZ}{N}. 
\ee
The only anomalies for a $\bZ_N$ symmetry that may be realized by Verlinde lines in diagonal RCFTs\footnote{These only admit quadratic refinements compatible with charge conjugation.} necessarily satisfy\cite{Lin:2021udi}:
\be \label{anomalconstr}
2k \in  N \bZ.
\ee
In particular, $\bZ_{\rm odd}$ must necessarily be non-anomalous, and for $\bZ_{\rm odd}$ only the order 2-subgroup in $\bZ_{2N}$ may carry a non-trivial anomaly. Matching these anomalies provides a selection rule for symmetry preserving flows.

Another strong constraint comes from the effective central charge theorem\cite{Castro-Alvaredo:2017udm}  asserting that the effective central charge $c_{\rm eff}  = c - 24 h_{\rm min}$ is monotonic:
\be 
c_{\rm eff}(\rm CFT_{UV}) \geq c_{\rm eff}(\rm CFT_{IR})
\ee
along $\mathcal{PT}$-symmetric RG flows, and generalizing the Zamolodchikov $c$-theorem\cite{Zamolodchikov:1986gt} to non-unitary flows.
Altogether, we will see that these conditions are sufficiently restrictive and provide an exclusive list of putative RG flows between $\cW_N$ RCFTs. 

\paragraph{\boldmath \texorpdfstring{$\cW$\textbf{-algebra minimal CFTs.}}{\textbf{W-algebra minimal CFTs.}}}
The diagonal (A-type) minimal models of the $\cW_N$ algebra, here denoted as $\cW_N(p,q)$, are a family of RCFTs  indexed by of pairs of coprime integers $p,q\geq N$ having central charge
\be
c\left[\cW_N(p,q)\right] =(N-1) \left(1-\frac{N(N+1)(p-q)^2 }{p q}\right).
\ee
They can be realized as the GKO cosets
\be \cW_{N}(p,q) = \frac{SU(N)_{\kappa-N}\times SU(N)_1}{SU(N)_{\kappa-N+1}}\comma\quad  \kappa = \frac{p}{q-p}\ee
and are non-unitary for $\abs{p-q}>1$.
$\cW_N(p,q)$  have a finite number $\frac{1}{N}\binom{p-1}{N-1}\binom{q-1}{N-1}$
of primary fields that can be parametrized by pairs of integrable highest weights $(\Lambda_L,\Lambda_R)$ of the algebra $\mathfrak{su}(N)$ at level $p-N$ and $q-N$ respectively, subjected to a simple current identification of order $N$. We describe in detail in  Appendix A how to produce a non-redundant set of primary fields. For the pair $(\Lambda_L,\Lambda_R)$ of Dynkin labels, we also employ the convenient graphical representation$$
W_{N}(p,q)\ni\phi = \Bigg( \overbrace{\ydiagram{6,5,2}}^{\leq p-N}\, ,\, \overbrace{\ydiagram{5,2}}^{\leq q-N} \Bigg\} < N \Bigg)
$$
in terms of pairs of Young diagrams.

In the following we will always refer, by abuse of notation, to a primary of the form  $\phi \in (\mydot, \Lambda_R)$ as  $\phi \in SU(N)_{q-N}$ and analogously for the left labels. This should not cause confusion as $p$ and $q$ are coprime.

Given a primary $(\Lambda_L,\Lambda_R)$, the modular $S$ matrices, the conformal weights $h(\phi)$ as well as their charge under the $\cW_N$ currents are known \cite{Lukyanov:1990tf}. We refer to Appendix A for more details.
Using this, we find
\be 
c_{\rm eff}\left[\cW_N(p,q)\right] =(N-1) \left(1-\frac{N(N+1)}{p q}\right).
\ee
\paragraph{\textbf{Virasoro RG flows.}}
To set up the stage and notation, let us briefly review the case of the simplest $\mathcal{W}_N$-algebra -- the Virasoro algebra $N=2$. Consider the set of fields labeled by the $k^{\rm th}$ Cartan powers of the adjoint representation of $SU(2)$ on the right and trivial on the left:
$$\phi_{[1,2k+1]} = \phi_{(0,2k)} = \Big(\mydot \comma \, \underbrace{\ydiagram{4}}_{2k}\,\Big). $$
They form a subcategory $(SU(2)_{q-2})_{\rm ad} \simeq SO(3)_{\lfloor \frac{q-2}{2}\rfloor}$
$$\left\{ (\mydot, \mydot) \comma (\mydot, \ydiagram{2})\comma \cdots\comma (\mydot,\overbrace{\ydiagram{6}}^{2 \lfloor \frac{q-2}{2}\rfloor}) \right\}$$
closed under fusion. The fusion category ${\rm Rep}[SU(2)_{p-2}]$ of Verlinde lines $\{\cL_\sigma\}$ corresponding to 
$$ 
\sigma \in \left\{ \Big(\mydot\comma\mydot\Big), \Big(\ydiagram{1}\comma\mydot\Big),\cdots ,\underbrace{\Big(\overbrace{{\ydiagram{5}}}^{p-2}\comma\mydot\Big)}_{\bZ_2 \, \text{simple current}} \right\}
$$
commutes with the adjoint subcategory  $(SU(2)_{q-2})_{\rm ad}$ as \eqref{simplecomm} is satisfied for $\cL_\sigma, \forall \sigma\in \Rep[SU(2)_{p-2}]$ and $\phi \in (SU(2)_{q-2})_{\rm ad}$. The top representative of this fusion category is the simple current ($\abs{d_{\mathcal{L}}} = 1$) generating the (possibly anomalous) $\mathbb{Z}_2$ symmetry of the minimal models.
Therefore, the non-invertible symmetry generated by $\Rep[SU(2)_{p-2}]$ is preserved by the adjoint-type deformation, and in particular that these flows  are always $\bZ_2$ symmetry preserving.
Based on this observation, in \cite{Nakayama:2024msv} (see also \cite{Ambrosino:2025yug}) matching the quantum dimensions of the  preserved lines lead to the following RG flows: 
\be \label{virflow}
\mathcal{W}_2(p,kp + i) \xrightarrow{\Big(\mydot \comma \,  \overbrace{\ydiagram{4}}^{2k} \Big)} \mathcal{W}_2(p,kp - i).
\ee
All the relevant fields in $(SU(2)_{kp+i-2})_{\rm ad}$ are dynamically generated, these being consistent with the preserved symmetries, with the field $(0, 2k)$ being the most relevant. 
This family may be continued also to to half-integer $k$. In this case, the flow is triggered by the symmetric power of the fundamental weight 
and the preserved subcategory of Verlinde line 
is the smaller\footnote{A note on the notation: by $\Rep[(SU(N)/\bZ_K)_k]$ we always mean the fusion category of the untwisted sector line.} $$
{\rm Rep}[SO(3)_{\lfloor \frac{p-2}{2} \rfloor}] \equiv {\rm Rep}[(PSU(2)_{p-2 })],$$ 
the neutral sub-category of $SU(2)_{p-2}$ under the central $\bZ_2$. 
Depending on the parity of $q$, these flow may or may not be $\bZ_2$-symmetry preserving. 
Let us also note that the field $(0,2k)$ is the unique field that, at leading order in $1/p$, is marginally relevant (irrelevant) in the UV (IR) theory\cite{Zamolodchikov:1987ti}:
$$ 
h^{\cW_2(p,kp\pm i)}_{[1,2k+1]} = 1 \mp \frac{i(k+1)}{k p}   + \mathcal{O} (1/p^2).
$$
We hope to come back to the Conformal perturbation theory expansion somewhere else.

\paragraph{\boldmath \texorpdfstring{\textbf{RG flows of  $\cW_3$ algebra. }}{\textbf{RG flows of  W3 algebra. }}}
Let us now consider the case of the $\cW_3$ algebra. 
Consider the primaries that form the adjoint orbit $(SU(3)_{q-3})_{\rm ad} \simeq (PSU(3)_{q-3})$ and are characterized by the objects in $SU(3)_{q-3}$ with trivial  triality ($N$-ality= \# boxes $= 0 \mod 3$): 
\begin{equation*}\label{closedfusion}
\begin{split} 
&(SU(3)_{k})_{\rm ad} = \{ (\lambda_1,\lambda_2) \in (SU(3)_{k})\,  | \, \, \lambda_1 + 2\lambda_2  \in 3 \bZ \}\\
&= \bigg\{ \mydot \comma \ydiagram{2,1}\comma \begin{array}{c}\ydiagram{3}\\[1ex]  \ydiagram{3,3} \end{array}  \comma  \ydiagram{4,2} \comma   \begin{array}{c}\ydiagram{5,1}\\[2ex]  \ydiagram{5,4} \end{array} \comma
\cdots \bigg\}
\end{split}
\end{equation*}
$(SU(3)_{q-3})_{\rm ad}$ contains the familiar Cartan powers of the adjoint representation\footnote{${\rm Adj}^k$ are only closed under fusion only once all the other representations with $N$-ality $ = 0 \mod 3$ are  included. }: ($k = 1,\cdots, \left\lfloor \frac{q-3}{2} \right\rfloor$)
\begin{equation*}\begin{split} 
{\rm \Adj}^k = \Big( (0,0), (k,k) \Big) 
= \bigg(\mydot, 
&\overbrace{\ydiagram{6,3}}^{2k}    \bigg) \comma \\ 
&\raisebox{8ex}{\vspace{-1.8cm}$\,\underbrace{\hphantom{\ytableaushort{~~~}}}_{k}$}
\end{split}
\end{equation*} 
\vspace{-1.1cm}

Note that complex conjugate representation must always appear in pairs and share the same conformal weight but opposite $W_3$ charge. In particular the fields $\Adj^k$ commute with $\cW_3$ zero mode.
Their conformal dimensions are strictly ordered by length of first row. 
These $(SU(3)_{q-3})_{\rm ad}$ have the crucial property (proven in Appendix B) of commuting with the full $${\rm Rep} [SU(3)_{p-3}]$$ fusion category of $(p-1)(p-2)$ simple objects given by the primaries with non-trivial left labels
\begin{align*}
\left\{ \begin{array}{c}\Big(\mydot\comma\mydot\Big), \Big(\ydiagram{1}\comma\mydot\Big),\cdots ,\underbrace{\Big(\overbrace{{\ydiagram{5}}}^{p-3}\comma\mydot\Big)}_{\bZ_3 \, \text{simple current}}\\\Big(\ydiagram{1,1}\comma\mydot\Big), \Big(\ydiagram{2,1}\comma\mydot\Big), \cdots ,\underbrace{\Big(\overbrace{{\ydiagram{5,5}}}^{p-3}\comma\mydot\Big)}_{\bZ_3 \, \text{simple current}} \end{array}\right\}
\end{align*}
that contains the 3 simple currents with right Dynkin labels $$\{(0,0), (p-3,0), (0,p-3) \}$$ generating the  $\bZ_3$ triality symmetry of the $\cW_3$ minimal models.

Then, the deformations triggered by $(SU(3)_{q-3})_{\rm ad}$ preserve the full ${\rm Rep}[SU(3)_{p-3}]$ symmetry as shown in Appendix B. 
As discussed around \eqref{anomalconstr}\cite{Lin:2021udi} Verlinde lines in a RCFT may only be a representation of a non-anomalous $\bZ_3$ symmetry.\footnote{One can indeed check that the monodromy  is always trivial: $3h( (p-3,0),(0,0))  =  (p-3) (q-3)\in \bZ$.}
\begin{figure*}
    \centering    \includegraphics[width=\linewidth]{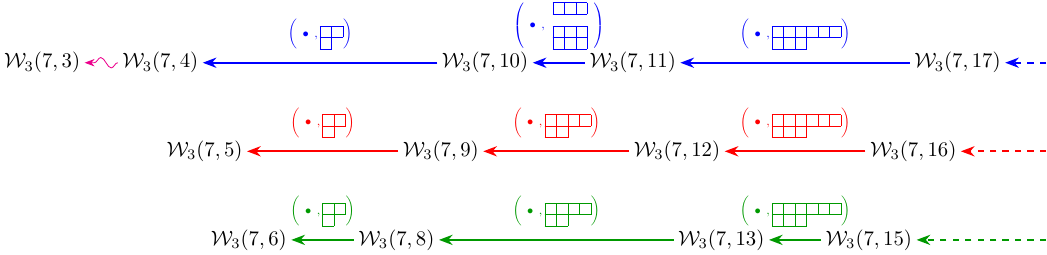}
    \caption{Flows between $\cW_3(7,q)$ minimal models. Cf.\ Appendix B for more details and examples for $3\leq q\leq 8$.}
    \label{fig:W3p7main}
\end{figure*}This implies that in the case of the $\cW_3$ algebra the global anomaly selection rule on the spin content of the Hilbert space is trivially satisfied. Yet, the quantum dimensions impose non-trivial constraints on the allowed flows. Indeed, it follows from direct computation that the set of quantum dimensions $\{d_\cL\}$ for  $\cL \in {\rm Rep}[SU(3)_{p-3}]$ in $\mathcal{W}_3(p,q)$ takes distinct values depending on $\pm q \mod p$. Therefore it is convenient to parametrize as in the Virasoro case $$  \{d_\cL\}[\cW_3(p,q)] \equiv \{d_\cL \}[ \mathcal{W}_3(p,kp + i)]  = \{d_{\cL}\} [p,k,i]. $$
Furthermore in Appendix B is proven:
$$\{d_\cL\}[p,k,i] = \{d_\cL\}[p,k,p-i]  = \{d_{\cL}\} [p,k,-i]  $$  
and otherwise take distinct values.
Then, imposing that the UV and IR quantum dimensions of the preserved lines match, as well as requiring that the effective central charge in the UV is greater than the one in the IR, leads to the following RG flows:
 \be\label{W3flows} \cW_3(p,\tfrac{k+1}{2} p + i) \xrightarrow{\Big(\mydot, \, \overbrace{  \hbox{\scalebox{0.7}{\ydiagram{4,2} }}}^{2k}\Big)} \cW_3(p,\tfrac{k+1}{2} p - i)
 \ee
for $k,2i \in \bZ$ such that  $3\leq kp + i \in \bZ$ and  coprime with $p$. 
Compared to Virasoro, where only the integer $k$ flows preserve the full $\Rep[SU(2)_{p-2}]$ symmetry, now also the half-integer $k$ preserve $\Rep[SU(3)_{p-3}]$; this is as a consequence of the absence of anomalies for the triality  symmetry.

Along the flows \eqref{W3flows} the perturbing field has been identified by being the most relevant among the adjoint subcategory, and it is generically $((0,0),(k,k))$. There are some special instances where $((0,0),(k,k))$ in $\cW_3(p,\frac{2k+1}{2}p +i)$ happens to be exactly marginal or slightly irrelevant; this  only happens for  $k$ odd and specific values of $q$ (some explicit examples are discussed in the appendix as well as in Figure \ref{fig:W3p7main}). In this case, the most relevant primary is always the one with $3k$ boxes.

Let us emphasize again that along the flows all the relevant fields in $SU(3)_{q-3}$ are  dynamically generated and their coupling must be fine tuned to reach the IR fixed point.  In appendix B we discuss in detail the cases with $p=3,4,5,6,7,8$ to give a concrete flavor of these flows.

Other generic primaries, obtained by (product of) Cartan powers of the fundamental weights
\begin{alignat*}{2} \omega_1^{k} &= \Big((0,0) , (k, 0) \Big)  &&=  \bigg(\mydot, 
\overbrace{\ydiagram{4}}^{k}    \bigg)  \notag\\ 
\omega_2^{k} &= \Big((0,0) , (0,k) \Big) &&=  \bigg(\mydot, 
\overbrace{\ydiagram{4,4}}^{k}    \bigg) 
\end{alignat*}
for $k\neq 0 \mod 3$  do not commute with the entire ${\rm Rep}[SU(3)_{p-3}]$ fusion category\footnote{Note that the only non trivial fusion sub-category of $SU(3)_k$ are  $PSU(3)_k$ and the bare $\bZ_3$ that is always enhanced in the RCFTs by its non-invertible completion.}, but only with its adjoint subcategory
$$ 
{\rm Rep}[PSU(3)_{p-3}] \simeq {\rm Rep}[(SU(3)_{p-3})_{\rm ad}]
$$
consisting of the lines in  ${\rm Rep}[(SU(3)_{p-3})]$ that have zero triality charge. Given that $\cW_3$ has a non-anomalous triality symmetry, no additional flows are
triggered by these classes of deformations. This does not imply that the flows triggered by such a deformation are trivial. Indeed, the existence of this large preserved symmetry category  ${\rm Rep}[PSU(3)_{p-3}]$ implies that the IR vacua of the gapped theory must define a module over it. We leave the discussion of gapped phases for future work \cite{Copetti:2024dcz,Copetti:2024rqj,  Chang:2018iay, Cordova:2024vsq,Cordova:2024iti,Cordova:2024goh,Cordova:2024iti}. 
 The list of deformations we have then provided is the complete assuming the preservation of the symmetry and having a generic level.

\paragraph{\boldmath \texorpdfstring{\textbf{RG flows of  $\cW_4$ algebra.}}{\textbf{RG flows of  W4 algebra.}}}
Let us now discuss the $\cW_4$ algebra before turning to the general case. This case is already much richer than $\cW_2$ or $\cW_3$. We will be more concise since we have already thoroughly explained the method when discussing $\cW_3$. The minimal models $\cW_4(p,q)$ enjoy the $\bZ_4$ symmetry generated by the 4 simple currents: \begin{align*} &\{ (0,0,0), (p-4,0,0) , (0,p-4,0) , (0,0,p-4) \}: \\  &\bZ_4 = \Bigg\{  \Big(\mydot,\mydot\Big), \Big(\scalebox{0.7}{\ydiagram{3}},\mydot\Big), \Big(\scalebox{0.7}{\ydiagram{3,3}},\mydot\Big) ,\Big( \, \scalebox{0.7}{\ydiagram{3,3,3}},\mydot\Big)\Bigg\}
\end{align*}
that is completed to the full modular fusion category
\begin{equation*} {\rm Rep }(SU(4)_{p-4}) \end{equation*}  
of the primaries with non-trivial left label.
The full ${\rm Rep }(SU(4)_{p-4})$ is preserved by the right primaries belonging to the adjoint subcategory and having vanishing $\bZ_4$ charge as proven in Appendix C:
$$ (SU(4)_{p-4})_{\rm ad} = \{\bs\lambda \in SU(4)_{p-4} | \lambda_1 + 2\lambda_2 + 3\lambda_3 \in 4\bZ\}.$$
Here ($\bs\lambda  = (\lambda_1,\lambda_2,\lambda_3)$). It is straightforward to match the anomalies from the UV to the IR, as we did above for $\cW_3$. This time we find that the set of quantum dimensions takes distinct values  $\pm q \mod 2p$ and satisfies
${d_\cL}[p,k,i] = {d_\cL}[p,k,-i]$. 
In particular, the flows with $k,i\in \frac{1}{2}\bZ$ do not preserve the full $SU(4)_{p-4}$,
\be\label{dodd}{d_{(\cL,\rm odd})}[p,k,p-i] = -{d_{(\cL,\rm odd})}[p,k,p-i]\ee
where by ${(\cL,\rm odd)}$ we indicate a representation with odd $N$-ality charge.
Hence, the flows that preserve the  full $\Rep[SU(4)_{p-4}]$ fusion category are
\be\label{W4flows}
\mathcal{W}_4(p,kp + i) \to\mathcal{W}_4(p,kp - i) \qquad k,i \in \bZ 
\ee 
exactly as in the Virasoro case. 
\begin{figure*}
    \centering
\includegraphics[width=\linewidth]{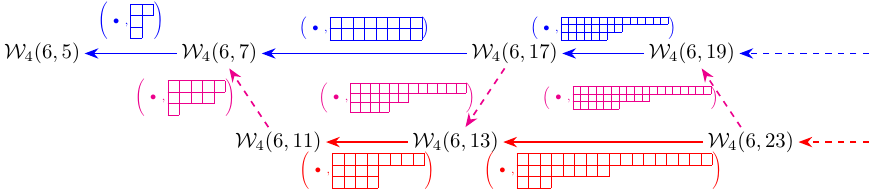}
    \caption{Flows between $\cW_4(6,q)$ minimal models. The flows in magenta (between top and bottom lines) are the one that preserve only $\Rep[(SU(4)/\bZ_2)_{q-4}]$ category. Cf.\ Appendix C for more details and detailed discussion $4\leq q\leq 8$}
    \label{fig:w4p6}
\end{figure*}

There are other interesting deformations corresponding to possible smaller fusion subcategories compatible with $\mathfrak{su}(4)_k$:
$$ \Rep[SU(4)_k] \qquad \Rep[(SU(4)/\bZ_2)_k] \qquad  \Rep[PSU(4)_k].$$
They correspond to the closed subcategories in $\Rep[\mathfrak{su}(4)_k]$ distinguised by their charge under the $\bZ_4$ center: 
$$\mathcal{C}_{SU(4)_k} = \cC_0 \oplus \cC_1 \oplus \cC_2 \oplus \cC_3 = \cC_{\rm even} \oplus \cC_{\rm odd}.$$
In addition to $ \cC_{SU(4)_k}\leftrightarrow\Rep[SU(4)_k]$ and $\cC_{0}\leftrightarrow\ \Rep [PSU(4)_k] \simeq (SU(4)_k)_{\rm ad}$, there is an additional  closed fusion category $\cC_{\rm even}\leftrightarrow\Rep [(SU(4)/\bZ_2)_k]$.
 
The deformations by right primaries $(SU(2)/\bZ_2)_{q-4}$ commute with exactly   $(SU(4)/\bZ_2)_{p-4}$ as shown in Appendix C. As such, these flows preserve only the (non-anomalous) $\bZ_2 \subset \bZ_4$ that is always there due to the anomaly being $k=0,2$. The corresponding flows are obtained by setting $k \in \frac{1}{2}+\bZ$. This mirrors exactly the analogous situation for $\cW_2$ where the $\omega_1^k$, commute only with $\Rep[PSU(2)_k]$ and generate the half-integer flows.  The generic object in $PSU(4)_{q-4}$ (powers of the fundamental weights) only commutes with the minimal adjoint subcategory $\Rep[PSU(4)_{p-4}]$, but this does not add any additional flows.

\paragraph{\textbf{The general case.}}
This discussion admits a straightforward generalization to a general $\cW_N$ algebra. In the minimal model $\cW_N(p,q)$ with $p>N+1$, the adjoint subcategory $(SU(N)_{q-N})_{\rm ad}$ commutes with the full $\Rep[SU(N)_{p-N}]$ subcategory of $\binom{k-1}{k-N}$ Verlinde lines. 
When $N$ is odd, the $\bZ_N$ symmetry is  always not anomalous, and one verifies that the quantum dimensions take the same values in $\pm q  \mod  p $. We therefore propose the $SU(2N+1)_{p-2N-1}$ symmetric RG flows:
$$ 
\cW_{2N+1}(p,\tfrac{2k+1}{2} p+i) \rightarrow \cW_{2N+1}(p,\tfrac{2k
+1}{2}p-i)
$$
Other deformations commuting with the subcategories $SU(N)/\bZ_k$ for a proper subgroup $\bZ_k$ of $\bZ_N$ do not induce any further flows. For even $N$ on the other hand, the anomaly of $\bZ_{2N}$ is $k =0, N$ and factors through its unique  index-2 subgroup
$$ 
0\to \bZ_N \to \bZ_{2N} \to \bZ_2\to 0$$
so that the anomaly is acting intrinsically only through its $\bZ_2$ subgroup \footnote{This is reflected by the $(-1)^k$ sign appearing in the quantum dimensions  for odd $N$-ality representations.}. It follows that, in this case, the quantum dimensions take values depending on the conjugacy classes $\pm q \mod 2p$. Based on this matching, we propose the $SU(2N)_{p-2N}$ preserving flows
$$ 
\cW_{2N}(p,k p+i) \rightarrow \cW_{2N}(p,k p-i)
$$
Those may be extended to $k \in \bZ_2$ allowing for the deformations preserving the smaller  $SU(2N)/\bZ_{N}$ fusion category generated by the Verlinde lines neutral under $\bZ_N$. The cases with $p = N, N+1$ are slightly special. In the first case, there is no $\bZ_N$ symmetry so no anomalies can be matched. When $p =N+1$, the $\Rep(SU(N)_1)\equiv \bZ_N$. For $\bZ_{\rm odd}$ there is no anomaly and all the contiguous flows are allowed, for $\bZ_{\rm even}$ the $\bZ_{2N}$-preserving flows are $\cW_{2N}(N+1, k+1) \to \cW_{2N}(N+1, k-1)$ and the the flows that preserve $\bZ_{N}$ are $\cW_{2N}(N+1, k+1) \to \cW_{2N}(N+1, k)$. \\
Another convenient parametrization for the same flows \eqref{WNflows} is
\be\begin{split}\label{WNflows2}
    \cW_{2N}(p,q) &\rightarrow \cW_{2N}(p,2kp-q)\comma\quad k \in \bZ \;\text{or} \;\bZ/2 \\
    \cW_{2N+1}(p,q) &\rightarrow \cW_{2N+1}(p,kp-q)\comma \quad k \in \bZ 
\end{split}
\ee
that may be replaced by the uniform  \be\cW_{N}(p,q) \rightarrow \cW_{N}(p,kp-q) \comma \qquad k \in \bZ \ee if we do not keep trace of the smaller preserved subcategory in the $\cW_{2N}$ case. Let us remark that this form might be suggestive of a $\cW_\infty$ origin: indeed it corresponds to
\be\label{Winfity} \cW_{\infty}(\lambda_2 , \lambda_3) \to \cW_{\infty}(-\lambda_2 - 2(k-1) \lambda_3, \lambda_3) \ee
at the level of the $\cW_\infty$ parameters\footnote{We thank Davide Gaiotto for discussions on this point.}. More details on the relation $(p,q)\leftrightarrow (\lambda_2,\lambda_3)$ and on $\cW_\infty$ algebra may be found in Appendix A. Notice that for the simplest RG flows with $k=1$, the $\lambda_j$ parameters in UV and IR are related in very simple way: one of them is kept fixed and the other one is reflected (independently of the rank $N$ of $\cW_N$ algebra!).

Let us conclude noting that the reason for such a straightforward generalization shall be traced back to the simplicity of the anomalies in diagonal RCFTs -- that even for $N>2$, due to the absence of anomalous subgroups of $\bZ_N$ that are not of index 2, all anomalies are of $\bZ_2$ type. 
Even though they do not provide any additional  RG flows, the deformations by primaries in $(SU(N)/\bZ_{M})_{q-N}$ commute with $\Rep[(SU(N)/\bZ_K)_{p-N}]$ for any proper subgroups of $\bZ_N$ -- $\bZ_M$ and $\bZ_K$ -- such that $\bZ_N/\bZ_M \simeq\bZ_K$. In particular, generic primaries in the right $SU(N)_{q-N}$ commute only with the adjoint subcategory $\Rep[SU(N)_{p-N}]$. Those may be relevant for the case of gapped flows.
The deformation that we have identified include and extend all the known integrable, both massless and massive, 
deformations of minimal models\cite{Vaisburd:1994hg}.
In Appendix D, the reader may found a detailed account of this. In addition, for special affine levels exceptional Frobenius algebras may be found.
\paragraph{\boldmath \texorpdfstring{ \textbf{Flows between $\cW_N$.}}{Flows between WN} }
As one deforms the CFT by a relevant primary, all the $\cW_N$ currents are broken. While $\cW_2$  re-emerges in the IR due to the conformal symmetry of the fixed point, the emergence of $\cW_{N>2}$ at the end point of the flow is more delicate.  

Let us take the example of the $\cW_3$ to illustrate this. 
The adjoint fields of the form $((0,0),(k,k))$ commute 
with the zero mode of the $\cW_3$ current, begin self-conjugate. Hence, both in the leading order conformal perturbation theory near the UV and the IR, the deformation is at least compatible with $\cW_3$ symmetry. Furthermore, given that the deformation preserve the $\bZ_3$ symmetry, the IR is naturally endowed with it enhancing the $\bZ_2$ symmetry of the Virasoro models. Yet, this is not sufficient to expect the $\cW_3$ symmetry to reemerge. 
Indeed, in the large space of possible flows obtained by tuning the relative couplings of the symmetry preserving relevant fields, the ones that do not change the rank of the $\cW$-algebra are very fine tuned ones, and lie  at very special points in the large space of couplings. Using the same techniques of this paper, one can also study flows between minimal models of $\cW$-algebras with distinct rank:
$\cW_N(p,q) \to \cW_{N'}(p',q')$. In this case, the preserved lines must flow to the IR with their own multiplicity: for $N> N'$ and for fixed quantum dimensions, there may be more lines in the UV having the same quantum dimension of a given IR line. As an example, consider the minimal models $\cW_{3}(5,q)$ for $q=\pm 1\mod 5 $ and  $\pm 2\mod 5$ the set of quantum dimension of $SU(3)_{2}$ coincide with ones in  $\Rep[SU(2)_3]$ of $\cW_{2}(5,q')$for  $q'=  \pm 2$ and $\pm 4 \mod 10 $ respectively. Yet,  for each pair of lines in $SU(2)_3$ there is a triality pair of lines in $SU(3)_2$ sharing the same dimension. 
Hence, in a putative flow between these models more than one simple line of the UV would flow to the same line of the IR. In this case, while the relevant field deforming the theory in the UV preserve the $\cW_3$ charge, the marginal operator controlling the near-IR theory breaks the $\cW_3$ zero mode  in general. 
We do not study this class of flows here in detail , but we hope to come back to this in the future. 

\paragraph{\textbf{Outlook.}}
In this letter we proposed the new RG flows between $\cW_N$ minimal models: 
$$\cW_{N}(p,q)\to \cW_{N}(p,kp-q)$$ 
The beautiful structure underlying these flows suggests many interesting future directions and generalizations in addition to the ones already mentioned.
Our proposed new flows 
are a perfect target for numerical exploration based e.g.\ on Hamiltonian or other numerical methods\cite{Lencses:2023evr,Delouche:2024tjf,Benedetti:2026tpa, Benedetti:2026tpa}, or possibly via $\epsilon$-expansion\cite{Klebanov:2022syt}, and should be tested thoroughly. 
Another interesting approach is using the RG interface techniques\cite{Gaiotto:2012np,Poghosyan:2023brb} that have been already employed for the unitary $W_2$ as well as for $\cW_3$ RG flows. Similarly, it would also be interesting to use these flows to derive novel Landau-Ginzburg descriptions along the lines of \cite{Klebanov:2022syt,Katsevich:2024jgq,Katsevich:2025ojk}.   Here we have restricted ourselves to the case of diagonal modular invariants, but it is important to extend this discussion also to the non-diagonal modular invariant minimal models. Some recent progress has been achieved in \cite{Tanaka:2025qou}, but in the $\cW_N$ much less is known explicitly\footnote{F.A.\ Acknowledges ongoing discussion with Connor Behan on the topic.}\cite{Frohlich:2009gb,Gannon:1992ty}. There the anomalies do not have to satisfy $k=0, N/2 \mod N$
and may play an important role. In \cite{Lin:2021udi}  constraints on $\bZ_N$ symmetry preserving operators in 2d CFT have been imposed using the modular bootstrap\cite{Collier:2016cls}.

Minimal models correspond to double truncations $Y_{N_1,N_2,N_3}\cap Y_{M_1,M_2,M_3}$ of the $\cW_{\infty}$ algebra\cite{Gaberdiel:2012uj,Prochazka:2015deb,Gaiotto:2017euk,Prochazka:2017qum,Prochazka:2018tlo, Prochazka:2024xyd}, of the special kind $  \cW_N(p,q) \simeq Y_{0,0,N}\cap Y_{p-N,q-N,0}$. Can the analysis of this paper be extended to  more general flows between corner vertex algebras $Y_{N_1,N_2,N_3}$?  Even more generally, we may hope to eventually extend these flows to the full non-rational $\cW_\infty$ (cfr.\ with the discussion around \eqref{Winfity}. 
Incidentaly, these new flows beg for an  interpretation in the context of supersymmetric field theories\cite{Gaiotto:2017euk}.
$\cW_N(p,q)$ are special cases of  GKO cosets of the form
\be \label{cosets}
\frac{SU(N)_\kappa \times SU(N)_\rho}{SU(N)_{\kappa+\rho}} \simeq \frac{U(\kappa+\rho)_N}{U(\kappa)_N \times U(\rho)_N}.
\ee
that arise as rational truncations of the more general Grassmanian VOAs \cite{Eberhardt:2020zgt}. 
The case $\rho =1$ and generic $N$ is the one  discussed in this letter. The $\rho=2, N=2$, corresponds is the ${\cal N}=1$ supersymmetric Virasoro case and is the subject of the recent investigation \cite{Gaberdiel:2026sfg} where analogous flows have been studied following the same strategy. 
 In addition, it could be interesting to study even the somewhat simpler gapless cosets $U({\kappa+\rho})_N/U(\kappa)_{N}$ that are truncations of rectangular $W^{(\rho)}_\infty$ algebras\cite{Creutzig:2018pts,Eberhardt:2019xmf}.
We are hopeful that our discussion may also be extended the more general class above, and we plan to report somewhere else. Furthermore, in \eqref{cosets} the affine $SU(N)$ algebra can be replaced by any other affine algebra. 

With the aim of uncover new integrable structures, it is important to study translational invariant defects\cite{Ambrosino:2025myh,Ambrosino:2025pjj} preserved by these new deformations. In the Virasoro case \cite{Ambrosino:2025myh,Ambrosino:2025pjj}, at least for fixed values  $p$ and $q$, the existence of infinitely many (non-local) conserved charges associated to translational invariant defects and preserved along flows by $(SU(2)_{p-2})_{\rm ad}$ was proven. It would be useful to extend this also to the $\cW_N$ case.
One may also hope that certain construction in 4d Chern-Simons \cite{Costello:2013zra} may be used to incorporate these new flows (as well as the old ones) coherently. In Appendix D, based on the discussion therein,  we propose a concrete target for possible integrable deformations:  the simplest representative from each of the possible symmetry preserving deformations: $(SU(N)/\bZ_K)_{p-N}$. 

We leave these and many more  exploration for future work.

\paragraph*{\textbf{Acknowledgments.}} 
FA thanks Davide Gaiotto, G\'erard M.\ T.\ Watts and Connor Behan, for numerous interesting discussions.
Research at Perimeter Institute is supported in part by the
Government of Canada through the Department of Innovation, Science and Economic Development Canada and
by the Province of Ontario through the Ministry of Colleges and Universities.
The research leading to these results has received support from the European Structural and Investment Funds and the Czech Ministry of Education, Youth and Sports (project No. \texttt{FORTE-CZ.02.01.01/00/22\_008/0004632}).\\
The authors gratefully acknowledge support from the Simons Center for Geometry and Physics, Stony Brook University and the organizers of the program \textit{Supersymmetric Quantum Field Theories, Vertex Operator Algebras, and Geometry} at which early stages of the research  for this paper have been performed.

\bibliography{references}

\appendix

\onecolumngrid
\newpage

\section*{Supplemental material}
\subsection{\boldmath \texorpdfstring{{Appendix A: $\cW_N$ minimal models}} {Appendix A: WN minimal models}}

\paragraph{\boldmath \textbf{$\cW_N$ algebras}}$\cW_N[c]$ algebras is a two-parametric family of vertex operator algebras labeled by integer rank parameter $N$ and central charge $c$ \cite{Bouwknegt:1992wg,Prochazka:2014gqa,Prochazka:2024xyd}. The algebra $\cW_N$ at generic $c$ is generated by stress-energy tensor $T(z)$ of dimension $2$ and independent primary fields of dimensions $3, 4, \ldots, N$: 
$$ \left\{T(z), W^{(3)}(z),\cdots , W^{N}(z) \right\}.$$
The simplest representative of the family is the Virasoro algebra having just the stress-energy tensor (and $N=2$).  For $N>2$, $\cW_N$ algebras are not linear. The simplest of those is the $\cW_3$ algebra generated by the stress-energy tensor $T(z)$ and a single spin $3$ current $W^{(3)}(z)$ \cite{Zamolodchikov:1985wn}. The OPE between $T(z)$ and the current of spin  $W^{(3)}(z)$ is dictated by the primary condition for $W^{(3)}(z)$,
$$ 
T(z) W^{(s)}(u) \sim \frac{s W^{(s)}(u)}{(z-u)^2} + \frac{\partial W^{(s)}(u)}{z-u} + reg.
$$
The OPE of $W^{(3)}(z)$ exhibits the non-linearity of the algebra
$$ 
W(z) W(u) \sim \frac{c(5c+22)}{144} \left[ \frac{1}{(z-u)^6} + \frac{\frac{6}{c} T(u)}{(z-u)^4} + \frac{\frac{3}{c}\partial T(u)}{(z-u)^3} +  \frac{1}{(z-u)^2} \left( \frac{9}{10c} \partial^2 T(u) + \frac{96}{c(5c+22)} {\color{blue}\Lambda(u)} \right) \right] + \cdots
$$
appearing through the composite quasi-primary operator $\Lambda(u) = (TT)(u) - \frac{3}{10} \partial^2 T(u)$. 
For higher spins the algebra becomes increasingly more complicated due to non-linearity. Apart from the non-linearity, there is an additional complication that contrary to the situation with $SU(N)$, there is no canonical embedding of $\cW_N$ as a subalgebra of $\cW_{N+1}$.

\paragraph{\textbf{Interpolating family}.}
For the reasons above it is very useful to consider a two-parametric family $\cW_\infty[\lambda,c]$ of algebras generated by stress-energy tensor and primary fields of dimension $3, 4, 5, \ldots$\cite{Blumenhagen:1994wg,Gaberdiel:2012uj,Prochazka:2014gqa,Linshaw:2017tvv}. The two parameters are the central charge $c$ and a rank-like parameter $\lambda$ that can take complex values. We can think of $\cW_\infty$ as an interpolating family of algebras in the following sense: for $c$ generic and $\lambda$ a positive integer $N$ the algebra $\cW_\infty$ develops an ideal $\mathcal{I}_N$ containing all primary generators of spin $N+1$ and higher and its quotient with respect to this ideal recovers $\cW_N$,
\be
\cW_\infty[\lambda = N,c] / \mathcal{I}_N = \cW_N[c]\period
\ee
Hence, while $\cW_N$ is not canonically embedded into $\cW_{N+1}$, upon passing to the larger $\cW_\infty$, we gain the advantage of a uniform description for all $\cW_N$. This comes at the cost of considering an algebra that is no longer finitely generated.
\paragraph{\textbf{Triality.}}
One of the advantages of working with the interpolating family $\cW_\infty$ is that it reveals a discrete triality symmetry of the algebra that underlies the whole representation theory. More concretely, for every fixed $c$ there are three values of parameter $\lambda$ satisfying conditions
\be
\label{centralcharge}
\frac{1}{\lambda_1} + \frac{1}{\lambda_2} + \frac{1}{\lambda_3} =0\comma \qquad c= (\lambda_1 -1)(\lambda_2 -1)(\lambda_3 -1)
\ee
such that
\be
\cW_\infty[\lambda_1,c] \simeq \cW_\infty[\lambda_2,c] \simeq \cW_\infty[\lambda_3,c].
\ee
In the following, we will equivalently parametrize $\cW_\infty$ algebras by triples of parameter $(\lambda_1,\lambda_2,\lambda_3)$ subjected to the constraint \eqref{centralcharge}.
\paragraph{\textbf{Truncations}.}
The truncation above reduces $\cW_\infty$ to $\cW_N$ by quotienting by the ideal $\cI_N$ is a very special truncation occurring when any of the $\lambda_j$ is a positive integer $N$. This however does not exhaust all of the possible truncations of the algebra. If the values of $(\lambda_1,\lambda_2,\lambda_3)$ satisfy the constraint
\be \label{Wtruncation}
\frac{N_1}{\lambda_1}+\frac{N_2}{\lambda_2}+\frac{N_3}{\lambda_3} = 1\comma\qquad N_i \in \bN,
\ee
there is a null state at level $(N_1+1)(N_2+1)(N_3+1)$ in the vacuum representation \cite{Prochazka:2014gqa,Prochazka:2017qum} and the quotient by the ideal $\mathcal{I}_{N_1,N_2,N_3}$ generated by this null state reduces to the corner vertex algebra
\be \cW_\infty[\lambda_1,\lambda_2,\lambda_3]/\mathcal{I}_{N_1,N_2,N_3} \simeq Y_{N_1,N_2,N_3}\comma \qquad \frac{N_1}{\lambda_1}+\frac{N_2}{\lambda_2}+\frac{N_3}{\lambda_3} = 1.
\ee
independently constructed in \cite{Gaiotto:2017euk}. The condition \eqref{Wtruncation} defines codimension one loci in the two-dimensional parameter space of $\cW_\infty$. 
Imposing simultaneously two independent conditions of the type \eqref{Wtruncation}, we find special codimension 2 points in parameter space
\be 
(\mathcal{W_\infty}/\cI_{N_1,N_2,N_3})/\cI_{M_1,M_2,M_3} \simeq \mathcal{W_\infty}/(\cI_{N_1,N_2,N_3}+\cI_{M_1,M_2,M_3})
\ee
that have rational central charge. Let us indicate, by an abuse of notation, the rational CFTs arising from this truncation as
\be 
Y_{N_1,N_2,N_3}\cap Y_{M_1,M_2,M_3} := \mathcal{W_\infty}/(\cI_{N_1,N_2,N_3}+\cI_{M_1,M_2,M_3}).
\ee
All $\cW_N$ minimal models arise from truncations of this kind. 
In particular, they are simultaneous truncations corresponding to $Y_{0,0,N}$ and $Y_{p-N,q-N,0}$,
\be 
\cW_N(p,q) = \mathcal{W_\infty}/(\cI_{0,0,N}+\cI_{p-N,q-N,0})
\ee
The parameters $p,q$ are related to $(\lambda_1,\lambda_2,\lambda_3)$ by solving the double truncation condition:
\be 
\lambda_1 =  \frac{N(p-q)}{q} \comma\quad \lambda_2 = \frac{N(q-p)}{p}\comma \quad \lambda_3 = N.
\ee
With this choice of parameters, the algebra has central charge
\be
c\left[\cW_N(p,q)\right] =(N-1) \left(1-\frac{N(N+1)(p-q)^2 }{p q}\right) = (\lambda_1-1)(\lambda_2-1)(\lambda_3-1)
\ee
as follows from \eqref{centralcharge}.

The diagonal $A$-type minimal models $\cW_N$ minimal models can be also realized via a GKO coset construction,
\be \cW_{N}(p,q) = \frac{SU(N)_{\kappa-N}\times SU(N)_1}{SU(N)_{\kappa-N+1}}\comma\qquad  \kappa = \frac{p}{q-p} \quad \text{or} \quad \kappa = \frac{q}{p-q}\ee
In this description, the action of triality is rather non-trivial, but the identification \be\kappa \leftrightarrow \kappa' = -1 -\kappa\ee implements exactly the symmetry $p\leftrightarrow q$ that would be rather mysterious otherwise at the level of the GKO coset. Indeed one may verify that
\be 
c({\rm GKO}) = c(SU(N)_{\kappa-N}) + c( SU(N)_1) - c(SU(N)_{\kappa-N+1})  \comma   \qquad
c\left(SU(N)_k\right) = \frac{k(N^2-1)}{k+N}
\ee
gives:
\be 
c({\rm GKO}) = \frac{(N-1)(N(N+1)- \kappa (1+\kappa))}{\kappa (1+\kappa)} =\eval_{\kappa = \frac{p}{q-p}} =  (N-1)\left(1- \frac{N(N+1)(p-q)^2}{p q} \right)
\ee
and that exhibits manifestly the symmetry $\kappa \leftrightarrow -1-\kappa$.
is the central charge of the Sugawara stress-energy tensor in affine Lie algebra $SU(N)_k$. The level $\kappa\in\bZ$ if and only if the model is unitary $\abs{p-q} = 1$ .

\paragraph{\textbf{Representations}}
Representation theory of the $\cW_\infty$ algebra may be nicely given exploiting the alternative description of the algebra in terms of the Yangian of affine $\mathfrak{gl}_1$, $\mathcal{Y}[\widehat{\mathfrak{gl}}_1]$ \cite{feigin2012quantum,Prochazka:2015deb}. The affine Yangian provides a nice description of large class of degenerate representations of $\cW_\infty$. The representations of this family are labeled by a triple of asymptotic Young diagrams and the states in the representation spaces correspond to plane partitions with given asymptotics. This family of degenerate primaries exhausts the primaries in the Kac table of $\cW_N$ minimal models.

Imposing a single truncation condition leading to $Y_{N_1,N_2,N_3}$ prohibits a box at coordinates $(N_1+1,N_2+1,N_3+1)$ which is exactly the highest null state that we are quotienting when passing from $\cW_\infty$ to $Y_{N_1,N_2,N_3}$ \cite{Bershtein:2018pcf}. If we impose two conditions of this kind, the space of plane partitions effectively becomes periodic \cite{Prochazka:2023zdb}. As a result of this, the degenerate primaries of $\cW_N(p,q)$ are labeled by pairs of Young diagrams $(\Lambda_L,\Lambda_R)$ with $N$ rows and $p-N$ or $q-N$ columns corresponding to asymptotics along the two directions. The periodicity imposes further identifications between the pairs of asymptotic Young diagram generated by the following three operations:
\begin{itemize}
 \item add a full column of length $N$ to $\Lambda_L$;
 \item add a full column of length $N$ to $\Lambda_R$;
 \item remove simultaneously a row of length $p-N$ from $\Lambda_L$ and $q-N$ from $\Lambda_R$.
\end{itemize}
The first two conditions let us choose the Young diagrams as labeling an irreducible representation of the $\mathfrak{su}(N)$ algebra. We can therefore label the primaries of $\cW_N(p,q)$ by pair  of integrable highest weights for
$(\Lambda_L,\Lambda_R)$  $SU(N)_{p-N}$ and  $SU(N)_{q-N}$ respectively, subjected to the \textit{diagonal identification} along the orbits generated by the moves 1-3. The orbit identification above may be readily implemented e.g.\ in \texttt{Mathematica}, to produce the pairs of independent primaries of $ \cW_N(p,q)$. Their number is given by remarkably simple formula
\be
\# \text{primaries} = \frac{1}{N} \binom{p-1}{N-1} \binom{q-1}{N-1} .
\ee
As an illustration, let us list the primaries for the minimal models $ \cW_{3}(5,6)$:
\begin{align}
\left\{ \begin{array}{c}
\Big(\mydot, \mydot\Big) \comma 
\Big( \vcenter{\hbox{\scalebox{0.6}{ \ydiagram{1}}}},\mydot \Big)\comma  
\Big( \vcenter{\hbox{\scalebox{0.6}{ \ydiagram{2}}}},\mydot\Big)\comma 
\Big( \vcenter{\hbox{\scalebox{0.6}{ \ydiagram{1,1}}}},\mydot\Big)\comma 
\Big( \vcenter{\hbox{\scalebox{0.6}{ \ydiagram{2,1}}}},\mydot\Big)\comma
\Big( \vcenter{\hbox{\scalebox{0.6}{ \ydiagram{2,2}}}},\mydot\Big)\\
\Big(\mydot, \vcenter{\hbox{\scalebox{0.6}{ \ydiagram{1}}}}\Big)\comma 
\Big(\mydot, \vcenter{\hbox{\scalebox{0.6}{ \ydiagram{2}}}}\Big)\comma
\Big(\mydot, \vcenter{\hbox{\scalebox{0.6}{ \ydiagram{1,1}}}}\Big)\comma
\Big(\mydot, \vcenter{\hbox{\scalebox{0.6}{ \ydiagram{2,1}}}}\Big)\comma 
\Big(\mydot, \vcenter{\hbox{\scalebox{0.6}{ \ydiagram{2,2}}}}\Big)\comma 
\Big(\mydot, \vcenter{\hbox{\scalebox{0.6}{ \ydiagram{3,1}}}}\Big)\comma 
\Big(\mydot, \vcenter{\hbox{\scalebox{0.6}{ \ydiagram{3,2}}}}\Big)\\
\Big(\vcenter{\hbox{\scalebox{0.6}{ \ydiagram{1}}}}, \vcenter{\hbox{\scalebox{0.6}{ \ydiagram{1}}}}\Big)\comma
\Big(\vcenter{\hbox{\scalebox{0.6}{ \ydiagram{1,1}}}}, \vcenter{\hbox{\scalebox{0.6}{ \ydiagram{1}}}}\Big)\comma
\Big(\vcenter{\hbox{\scalebox{0.6}{ \ydiagram{2,1}}}}, \vcenter{\hbox{\scalebox{0.6}{ \ydiagram{1}}}}\Big)\comma
\Big(\vcenter{\hbox{\scalebox{0.6}{ \ydiagram{1}}}}, \vcenter{\hbox{\scalebox{0.6}{ \ydiagram{2}}}}\Big)\comma
\Big(\vcenter{\hbox{\scalebox{0.6}{ \ydiagram{1,1}}}}, \vcenter{\hbox{\scalebox{0.6}{ \ydiagram{2}}}}\Big)\comma
\Big(\vcenter{\hbox{\scalebox{0.6}{ \ydiagram{2,1}}}}, \vcenter{\hbox{\scalebox{0.6}{ \ydiagram{2}}}}\Big)\comma
\Big(\vcenter{\hbox{\scalebox{0.6}{ \ydiagram{1}}}}, \vcenter{\hbox{\scalebox{0.6}{ \ydiagram{2,1}}}}\Big)\comma
\end{array}\right\}.
\end{align}
There are indeed exactly $\frac{1}{3}\binom{4}{2}\binom{5}{2}= 20$ of these. Among these, the reader may identify the two distinguished sub-fusion categories with the fusion rings of $SU(N)_{p-N}$ and $SU(N)_{q-N}$.

Given a primary, the conformal weight is given by the formula
\be
h(\Lambda_L, \Lambda_R) = 
\frac{\lVert q(\Lambda_L + \rho) - p(\Lambda_R  + \rho) \rVert^2 - (q - p)^2 \lVert \rho \rVert^2}{2pq}.
\ee
where $\rho$ is the Weyl vector and we identify $\Lambda_L$ and $\Lambda_R$ with the weights of $SU(N)$ \cite{Bouwknegt:1992wg, Prochazka:2015deb}.

\paragraph{\textbf{Labeling of primaries}}
Considering $\Lambda_L$ and $\Lambda_R$ as Young diagrams of at most $N-1$ rows and $p-N$ or $q-N$ columns, we have the restrictions
\be
p-N \geq \Lambda_{L,1} \geq \Lambda_{L,2} \geq \ldots \geq \Lambda_{L,N-1} \qquad \text{and} \qquad q-N \geq \Lambda_{R,1} \geq \Lambda_{R,2} \geq \ldots \geq \Lambda_{R,N-1}.
\ee
The corresponding Dynkin labels are
\be
a_j = \Lambda_{L,j} - \Lambda_{L,j+1}, \qquad b_j = \Lambda_{R,j} - \Lambda_{R,j+1}
\ee
and are subjected to
\be
\sum_{j=1}^{N-1} a_j \leq p-N \comma \qquad  \sum_{j=1}^{N-1} b_i  \leq q-N.
\ee
In other words, the Dynkin labels express the weight vectors associated to $\Lambda_L$ and $\Lambda_R$ as linear combinations of the fundamental weights
\be
\Lambda_L \rightarrow \sum_j a_j \omega_j, \qquad \Lambda_R  \rightarrow \sum_j b_j \omega_j
\ee
and the Weyl vector is simply the sum of fundamental weights of $SU(N)$,
\be
\rho = \sum_j \omega_j.
\ee
In labeling the primaries in 2d CFT the literature, one often employs the Kac labels which are just the Dynkin labels shifted by one,
\be 
r_i = a_i + 1 \comma \qquad s_i = b_i + 1.
\ee
We therefore have three different ways of labeling the primaries, by Dynkin labels
\be
((a_1,\ldots,a_{N-1}),(b_1,\ldots,b_{N-1})),
\ee
by Kac labels
\be[r, s]=  [r_1,\cdots, r_{N-1} ; s_1,\cdots, s_{N-1}] : = 
\left[ \begin{array}{cc}
r_1 & s_1 \\
\vdots & \vdots\\
r_{N-1} & s_{N-1}
\end{array}\right]
\ee
or by Young diagrams
\be
\Bigg( \overbrace{\ydiagram{6,5,2}}^{\leq p-N}\, ,\, \overbrace{\ydiagram{5,2}}^{\leq q-N} \Bigg\} < N \Bigg)
\ee
where $\Lambda_{L,j}$ and $\Lambda_{R,j}$ are the lengths of the rows. We find the labels by Dynkin labels and Young diagrams to be a more convenient representation for the primary fields and therefore we will not employ Kac notation in this paper except in Appendix D when comparing to the literature.

\paragraph{\textbf{Modular S-matrix.}}
The modular S-matrix for the minimal models of the $\cW_N$ algebra is known, and takes the form \cite[p.~71]{Lukyanov:1990tf}
\be\begin{split}
 S_{(\Lambda_L,\Lambda_R),(\Lambda'_L,\Lambda'_R)}
 & =  \frac{1}{\sqrt{N} (pq)^{(N-1)/2}}
\exp \left[ 2\pi i (\Lambda_L+\rho)\cdot(\Lambda'_R+\rho) + 2\pi i (\Lambda'_L+\rho)\cdot(\Lambda_R+\rho) \right] \times \\
 \times &\sum_{w \in \mathcal{S}_N} ({\rm sgn}\, w) \cdot \exp \left[ -\frac{2\pi i q}{p} (\Lambda_L+\rho) \cdot w(\Lambda'_L+\rho) \right]  \sum_{w \in \mathcal{S}_N} ({\rm sgn}\, w) \cdot \exp \left[ -\frac{2\pi i p}{q} (\Lambda_R+\rho) \cdot w(\Lambda'_R+\rho) \right]
\end{split}
\ee
\paragraph{\textbf{Fusion rules and Verlinde formula.}}
The primary fields form the fusion ring:
\be 
\left[\phi_{\rho} \right] \times \left[\phi_{\sigma} \right] = \sum_{\kappa} N^{\kappa}_{\rho\sigma} \,\left[\phi_{\kappa}\right]  
\ee
where the fusion coefficients $N^{\kappa}_{\rho\sigma}$ are given in terms of the modular S-matrix  $S_{\rho\sigma}$ via the Verlinde formula:
\be
    N^\kappa_{\rho\sigma} = \sum_\lambda \frac{S_{\rho\lambda}\,S_{\sigma\lambda} S_{\kappa\lambda}}{S_{0\lambda}}
\ee
Unlike in the Virasoro case where the fusion multiplicities $N^\kappa_{\rho\sigma}$ are either $0$ or $1$, in the case of $\cW_N$ the multiplicities are non-negative integers possibly greater than one according to the $SU(N)_k$ fusion rules. As an example  consider the following fusion in $\cW_3(p,q \geq 7)$:
$$ 
\Big(\mydot, \vcenter{\hbox{\scalebox{0.7}{\ydiagram{2,1}}}}\Big) \times  \Big(\mydot, \vcenter{\hbox{\scalebox{0.7}{\ydiagram{2,1}}}}\Big)  = \Big(\mydot, \mydot\Big) + \Big(\mydot, \vcenter{\hbox{\scalebox{0.7}{\ydiagram{2,1}}}}\Big) + \Big(\mydot, \vcenter{\hbox{\scalebox{0.7}{\ydiagram{2,1}}}}\Big) +  \Big(\mydot, \vcenter{\hbox{\scalebox{0.7}{\ydiagram{3}}}}\Big)+  \Big(\mydot, \vcenter{\hbox{\scalebox{0.7}{\ydiagram{3,3}}}}\Big) +  \Big(\mydot, \vcenter{\hbox{\scalebox{0.7}{\ydiagram{4,2}}}}\Big)
$$
where $N_{{\rm Adj,  Adj}}^{\rm Adj} = 2$ whenever $q\geq 5$. 
\paragraph{Verlinde lines.}
The topological Verlinde line operators, are in one-to-one correspondence with the primary fields $\cL_\rho \equiv \cL_{(r,s)}$. They satisfy the same fusion ring as the primary fields:
\be 
\cL_{\rho} \times \cL_{\sigma} = \sum_{\lambda} N_{\rho\sigma}^{\lambda}\, \cL_{\lambda}  \period
\ee
Let us denote $\phi_\rho \ket{0} = \ket{\phi_\rho}$  a state on the cylinder. The action of a Verlinde line on a primary field is given by:
\begin{align}
    \cL_{\rho} \ket{\phi_{\sigma}} &= 
    \begin{tikzpicture}[baseline={(0,-0.5ex)}]
        \draw[ thick,red] (0,0) circle [radius=0.6cm];
        \draw[->,red,thick] (0.6,0) arc [start angle=0, end angle=120, radius=0.6cm];
        \fill (0,0) circle [radius=1pt];
            \node at (0.2, -0.2) {$\phi_{\sigma}$};
         \node at (-0.9, -0) {$ \color{red} \cL_{\rho}$};
    \end{tikzpicture} = \frac{S_{\rho\sigma}}{S_{0\sigma}} \ket{\phi_\sigma} \period
\end{align}
The action of a Verlinde line on a primary field, should be thought as generating the Ward identity for the discrete symmetry associated to a given topological line. 
The vacuum expectation value of a Verlinde line on the cylinder is known as \textit{quantum dimension}:\be 
\langle \cL_\rho\rangle= \bra{0}\cL_\rho \ket{0}  = 
 \begin{tikzpicture}[baseline={(0,-0.5ex)}]
        \draw[ thick,red] (0,0) circle [radius=0.6cm];
        \draw[->,red,thick] (0.6,0) arc [start angle=0, end angle=120, radius=0.6cm];
         \node at (-0.9, -0) {$ \color{red} \cL_{\rho}$};
    \end{tikzpicture} = d_\rho \period
\ee
Quantum dimension furnish a one-dimensional representation of the fusion algebra: 
\be 
d_{\rho} \cdot d_{\sigma} = \sum_{\lambda} N_{\rho \sigma}^{\lambda} d_{\lambda}
\ee
The reader may verify that all the quantum dimension reported in the examples in Appendices B and C are indeed consistent with the $SU(N)_k$ fusion rules.

\paragraph{\textbf{Effective central charge.}}

For $\mathcal{PT}$ symmetric flows, a generalization of the usual  $c$-theorem holds upon considering the effective central charge:
\begin{equation}
c_{\rm eff} \equiv c - 24 h_{\rm min}
\end{equation}
where $h_{\rm min}<0$ is the conformal weight of the lowest lying primary (for non-unitary CFTs this conformal weight is negative while for unitary CFTs it vanishes). This is the quantity entering the Cardy formula and controlling the asymptotic growth of the states. For $\cW_N(p,q)$ minimal models, there is a simple formula for the effective central charge:
\begin{equation}
c_{\rm {eff}} = (N-1)\left(1 - N(N+1)\frac{1}{pq}\right).
\end{equation}
It is valid for both unitary and non-unitary minimal models and for the unitary minimal models we can see this explicitly using the relation $p = q \pm 1$. Note also that for $N=2$ (Virasoro algebra), one indeed recovers the well known formula
\be
c_{\rm eff} = 1- 6/pq.
\ee

\paragraph{\textbf{Interesting generalizations.}}
Let us mention here some direct generalization that may be interesting for future applications. The GKO cosets giving the minimal models are a special case of a much broader class of coests CFTs: for any given affine algebra $\mathfrak{g}$ the GKO construction produces the rational cosets CFTs:
\be \frac{\mathfrak{g}_{\kappa-h^\vee} \times \mathfrak{g}_\rho}{\mathfrak{g}_{\kappa+\rho-h^\vee}}\ee
    the case $\mathfrak{g} = \mathfrak{su}(N)$ and $\rho = 1$ gives $\cW_N(p,q)$. $\rho=2$ corresponds to the supersymmetric minimal models, while higher integer values of $\rho$ give $\bZ_\rho$ graded algebras. Many of the considerations made in this paper should equally apply to these more general cases.

\subsection{\boldmath \texorpdfstring{Appendix B:  Detailed analysis for $\mathcal{W}_3$}{Appendix B:  Detailed analysis for W3}}
\subsubsection{\textbf{General proofs}}
Consider the minimal models $\cW_3(p, \frac{k+1}{2}p +i)$.

\paragraph{\textbf{Commutation condition.}}
Consider a generic right primary $\phi\in SU(N)_{q-1}$:
\be\begin{split} 
\label{phiprimary}
\phi = (\mydot, [k_1,k_2]) = \Big( (0,0), (k_1-k_2,k_2) \Big) 
= \bigg(\mydot, 
&\overbrace{\ydiagram{7,4}}^{k_1\leq p-3}    \bigg) \comma \\ 
&\raisebox{10ex}{\vspace{-2cm}$\,\underbrace{\hphantom{\ytableaushort{~~~~}}}_{k_2\leq k_1}$}
\end{split}
\ee \\[-1cm]
 and a line $\cL_\sigma \in \Rep[SU(3)_{p-3}]$:
\be\begin{split} 
\sigma = ( [l_1,l_2],\mydot ) = \Big( (l_1-l_2,l_2),(0,0) \Big) 
= \bigg(
&\overbrace{\ydiagram{7,4}}^{l_1\leq p-3} ,\mydot   \bigg) . \\ 
&\raisebox{10ex}{\vspace{-2cm}$\,\underbrace{\hphantom{\ytableaushort{~~~~}}}_{l_2\leq l_1}$}
\end{split}
\ee \\[-1cm]
Computing explicitly the commutation condition \eqref{simplecomm} in the minimal model $\cW_3(p,q)$ one finds: ($t = q/p$) 
\be 
d_{\cL_\sigma} - \frac{S_{\sigma \phi}}{S_{0 \phi}} = \left[\frac{\sin({\pi  (l_1- l_2+1) t}) \sin ({\pi  (l_1+2) t}) \sin ({\pi  (l_2+1)t})}{i(-1)^{(k_1+k_2)(l_1+l_2)} \sin ^3({\pi  t}) \cos ({\pi  t})}\right] {\color{blue}\sin (\frac{\pi}{3}  (2k_1 l_1 - k_2 l_1 - k_1 l_2 + 2 k_2 l_2 ) ) }
\ee
The factor in parentheses never vanishes given that $l_1, l_2, l_1-l_2 \leq p-3$. Instead, the term in blue color $\sin(\frac{\pi}{3}\#)$ vanishes  if and only if 
$$ 
(2k_1 l_1 - k_2 l_1 - k_1 l_2 + 2 k_2 l_2 ) \,  {\rm mod}\, 3  =  (l_1 + l_2) (k_1 + k_2) {\rm mod} \,3  = 0 
$$
Then,
\begin{itemize}
    \item if $k_1 + k_2 \in 3\bZ$, i.e.\ for $\phi \in (SU(3)_{q-3})_{\rm ad}$, $\Rep[SU(3)_{p-3}]$ is preserved
    \item otherwise, only $\Rep[PSU(3)_{p-3}]$ commutes with a generic deformation in $SU(3)_{q-3}$
\end{itemize}
as claimed in the main text. 

\paragraph{\textbf{Matching of quantum dimensions.}}
The quantum dimension of a line $  SU(3)_{p-2}\ni \cL = ((0,0), (l_1-l_2,l_2))$ is 
\be 
d_\cL[\cW_3(p,q)] =  \frac{1}{ \sin ^2(\pi  t) \cos (2 \pi  t)} \sin \left(\pi  \left(l_1+2\right) t\right) \sin \left(\pi  \left(l_1-l_2+1\right) t\right) \sin \left(\pi  \left(l_2+1\right) t\right).
\ee
Then, setting $q = k p \pm i$, after many sign cancellations
\be 
    d_\cL[\cW_3(p,q)] = \frac{1} {\sin ^2({\pi  i}/p) \cos ({2 \pi  i}/{p})} \sin ({\pi  i (l_1+2)}/{p}) \sin ({\pi  i (l_1-l_2+1)}/{p}) \sin ({\pi  i (l_2+1)}/{p})
\ee
and therefore $d_\cL[\cW_3(p,kp+i)] = d_\cL[\cW_3(p,kp-i)]$ as claimed in the main text. 
Analogously for the half-integer case $q = \frac{2k+1}{2}\pm i$ one has:
\be \begin{split}
d_\cL[\cW_3(p,q)] &=\csc  ( \pm\tfrac{2 \pi  i}{p} ) \sec ^2 \left(\pm \tfrac{\pi  i}{p} \right) \sin  (\pm\pi  i  \tfrac{(l_1+2 )}{p}+ \tfrac{\pi  l_1}{2} ) \cos  (\pm\pi  i  \tfrac{(l_1-l_2+1 )}{p} + \tfrac{1}{2} \pi   (l_1-l_2 ) ) \cos  (\pm\pi  i  \tfrac{(l_2+1 )}{p} + \tfrac{\pi  l_2}{2} )\\
&=\csc  ( \tfrac{2 \pi  i}{p} ) \sec ^2 \left( \tfrac{\pi  i}{p} \right) \sin  (\pm\pi  i  \tfrac{(l_1+2 )}{p}+ \tfrac{\pi  l_1}{2} ) \cos  (\pm\pi  i  \tfrac{(l_1-l_2+1 )}{p} + \tfrac{1}{2} \pi   (l_1-l_2 ) ) \cos  (\pm\pi  i  \tfrac{(l_2+1 )}{p} + \tfrac{\pi  l_2}{2} )
\end{split}
\ee
This does not depend on $i$ after using trivial trigonometric identities.
Therefore, as claimed in the text,
$$ 
d_\cL[\cW_3(p, \tfrac{k+1}{2}p+i)] = d_\cL[\cW_3(p, \tfrac{k+1}{2}p-i)].
$$

\subsubsection{\boldmath  \texorpdfstring{\textbf{Detailed illustration for} $p \leq 8$}{\textbf{Detailed illustration for} p < 8}}

\paragraph*{\boldmath $\mathcal{W}_3(3 , q )$} The case with $p=3$ is trivial as the left subcategory is $SU(3)_{p-3}$. This is always true for $\cW_N(N,q)$ and indeed the analogous phenomenon was already pointed out in \cite{Nakayama:2024msv}. In 
the case of $\cW_2$ the continuation to $\cW_2(2,2k+i)\to\cW_2(p,2k-i)$  with $k\in \bZ/2$ does indeed reproduces the flows found in \cite{Lencses:2023evr}. 
\paragraph*{\boldmath $\mathcal{W}_3(4 , q )$}
Let us consider the case of $p=4$, $\cW_3(4, 4k +i )$. In this case, the preserved subcategory \be{\rm Rep}[SU(3)_1] = \Bigg\{\(\mydot, \mydot \) \comma \left(\ydiagram{1}, \mydot \right) \comma \left(\ydiagram{1,1}, \mydot \)\Bigg\}
\ee
consists only of the simple currents generating the triality symmetry and having all quantum dimensions $d_\cL = 1$ (the only consistent solution, or equivalently no anomalies in $\bZ_3$ symmetry of diagonal RCFTs). This is preserved by the adjoint subcategory ${\rm Rep }[SU(3)_{q-3}]$. Then,  all the flows the flows allowed by symmetry-matching are:
\begin{align*}
\includegraphics[width=\linewidth]{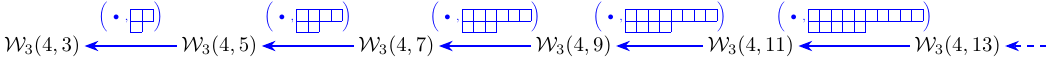}
\end{align*}

\paragraph{\boldmath $\mathcal{W}_3(5 ,q)$}
\begin{table}[]
    \centering
    \begin{tabular}
    {|c||c|c|c|c|c|c|}
     \hline
 \multicolumn{7}{|c|}{q-dimensions  in $\cW_3(5,q)$ of  ${\rm Rep}[SU(3)_2]$} \\
\hline   & $\left(\mydot, \mydot \right)$ & $\left(\ydiagram{1}, \mydot \right)$ &$\left(\ydiagram{1,1}, \mydot \right)$  &$\left(\ydiagram{2}, \mydot \right)$  & $\left(\ydiagram{2,1}, \mydot \right)$  & $\left(\ydiagram{2,2}, \mydot \right)$ \\
     \hline
        $q = 1,4 \mod 5$ & 1  &$\varphi$ & $\varphi$&1 & $\varphi$ &1\\
        $q = 2,3 \mod 5$ & 1  &$-1/\varphi$ & $-1/\varphi$&1 & $-1/\varphi$ &1        \\
        \hline
    \end{tabular}
    \caption{Quantum dimension of the preserved subcategory ${\rm Rep}[SU(3)_2]$ in the minimal models $\cW_3(5,q)$. $\varphi = \frac{1+\sqrt{5}}{2}$ is the golden ratio } 
    \label{tab:W3p5}
\end{table} 
In this case, the preserved subcategory is ${\rm Rep}[SU(3)_2]$. The quantum dimensions of the preserved subcategory in $\mathcal{M}(5,q)$ depend on $q \mod 5$ as in reported in Table \ref{tab:W3p5} and the resulting flows are:
\begin{align*}
\includegraphics[width=\linewidth]{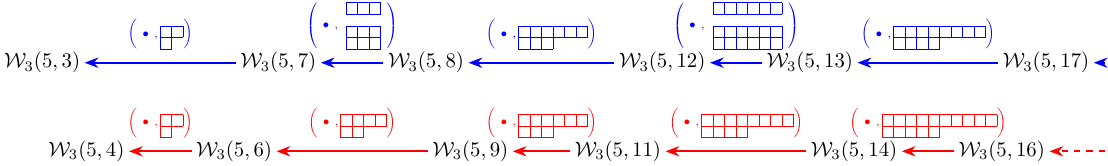}
\end{align*}

\paragraph{\boldmath $\cW_3(6, q)$}
\begin{table}[]
    \centering
    \begin{tabular}
    {|c||c|c|c|}
     \hline
 \multicolumn{4}{|c|}{q-dimensions  in $\cW_3(6,q)$ of  ${\rm Rep}[SU(3)_3]$} \\
\hline   & \makecell[c]{$\left(\mydot, \mydot \right)$ \\[1ex] $\left(\ydiagram{3}, \mydot \right)$\\[1ex] $\left(\ydiagram{3,3}, \mydot \right)$}
 & \makecell[c]{$\left(\ydiagram{1}, \mydot \right)$ \\[1ex] $\left(\ydiagram{2}, \mydot \right)$\\[1ex] $\left(\ydiagram{4,1}, \mydot \right)$}  \makecell[c]{$\left(\ydiagram{1,1}, \mydot \right)$ \\[1ex] $\left(\ydiagram{2,2}, \mydot \right)$\\[1ex] $\left(\ydiagram{4,3}, \mydot \right)$} & $\left(\ydiagram{2,1}, \mydot \right)$ \\
     \hline     
        $q = 1,5 \mod 6$ & 1  &$2$ & $3$       \\
        \hline
    \end{tabular}
    \caption{Quantum dimension of the preserved subcategory ${\rm Rep}[SU(3)_3]$ in the minimal models $\cW_3(6,q)$. } 
    \label{tab:W3p6}
\end{table}
In this case, the preserved subcategory is ${\rm Rep}(SU(3)_3)$. For $\mathcal{W}_3(6,q)$ the only coprime $q$ are $q = 1, 5 \mod 6$. The quantum dimensions of the ${\rm Rep}[SU(3)_3]$ fusion category for $q = 1,5$ are reported in Table \ref{tab:W3p6}. 
The resulting flows are: 
\begin{align*}
\includegraphics[width=\linewidth]{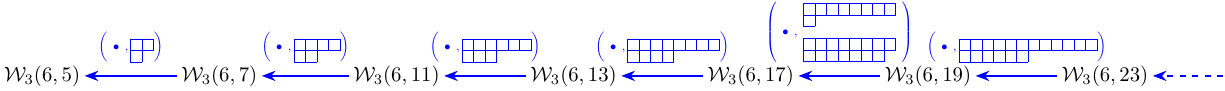}
\end{align*}

\paragraph{\boldmath $\cW_3(7, q)$}
\begin{table}[]
    \centering
    \begin{tabular}
    {|c||c|c|c|c|}
     \hline
 \multicolumn{5}{|c|}{q-dimensions  in $\cW_3(7,q)$ of  ${\rm Rep}[SU(3)_4]$} \\
\hline   & \makecell[c]{$\left(\mydot, \mydot \right)$ \\[1ex] $\left(\ydiagram{4}, \mydot \right)$\\[1ex] $\left(\ydiagram{4,4}, \mydot \right)$}
 & \makecell[c]{$\left(\ydiagram{1}, \mydot \right)$ \\[1ex] $\left(\ydiagram{3}, \mydot \right)$\\[1ex] $\left(\ydiagram{4,1}, \mydot \right)$}  \makecell[c]{$\left(\ydiagram{1,1}, \mydot \right)$ \\[1ex] $\left(\ydiagram{3,3}, \mydot \right)$\\[1ex] $\left(\ydiagram{4,3}, \mydot \right)$} &\makecell[c]{$\left(\ydiagram{2}, \mydot \right)$ \\[1ex] $\left(\ydiagram{2,2}, \mydot \right)$\\[1ex] $\left(\ydiagram{4,2}, \mydot \right)$} &\makecell[c]{$\left(\ydiagram{2,1}, \mydot \right)$ \\[1ex] $\left(\ydiagram{3,1}, \mydot \right)$\\[1ex] $\left(\ydiagram{3,2}, \mydot \right)$}  \\
     \hline
        $q = 1,6 \mod 7$ & 1  &$r_1$ & $s_1$&$t_1$\\
        $q = 2,5 \mod 7$ & 1  &$r_2$ & $s_2$&$t_2$        \\
        $q = 3,4 \mod 7$ & 1  &$r_3$ & $s_3$&$t_3$        \\
        \hline
    \end{tabular}
    \caption{Quantum dimension of the preserved subcategory ${\rm Rep}[SU(3)_4]$ in the minimal models $\cW_3(7,q)$. The values of $r_i,s_i,t_i$ are reported in \eqref{thirdroots}.} 
    \label{tab:W3p7}
\end{table}

The preserved category is ${\rm Rep}[SU(3)_4]$. The quantum dimensions of the objects in the minimal models $\cW_3(7,q)$ are listed in Table \ref{tab:W3p7}. In the table, $r_i,s_i,t_i$ are the roots of the following monic polynomials: 
\begin{alignat}{4}
p(r)&=r^3 - 2r^2 - r +1:  \qquad &&r_1 \simeq 2.24698\comma\quad  &&r_2 \simeq 0.554958\comma\quad &&r_3 \simeq -0.801938\nonumber\\
\label{thirdroots}
p(s)&=s^3 - 4s^2 + 3s +1:  \qquad &&s_1 \simeq 2.80194\comma\quad  &&s_2 = -0.24698\comma\quad &&s_3 = 1.44504\\
p(t)&=t^3 - 3t^2 - 4t -1:   \qquad &&t_1 \simeq 4.04892\comma\quad  &&t_2 = -0.692021\comma\quad &&t_3 = -0.356896\nonumber
\end{alignat}
These follows from consistency with the fusion rules. 
The flows are as follows:
\begin{align}
\includegraphics[width=\linewidth]{W3p7.pdf}
\end{align}

\paragraph{\boldmath $\cW_3(8, q)$}
As the last example we discuss $\cW_3(8, q)$. This is the first model in $\cW_3$ with $p$ even that allows for a non-trivial flow structure. Indeed the quantum dimension $\mod 8$ are split in 2 classes as shown in Table \ref{tab:W3p8}. The resulting flows are:
\begin{table}[]
    \centering
    \begin{tabular}
    {|c||c|c|c|c|c|}
     \hline
 \multicolumn{6}{|c|}{q-dimensions  in $\cW_3(8,q)$ of  ${\rm Rep}[SU(3)_5]$} \\
\hline   &\makecell[c]{$\left(\mydot, \mydot \right)$ \\[1ex] $\left(\ydiagram{5}, \mydot \right)$
}
 & \makecell[c]{ $\left(\ydiagram{1}, \mydot \right)$ \\[1ex] $\left(\ydiagram{4}, \mydot \right)$\\[1ex] $\left(\ydiagram{5,1}, \mydot \right)$} &  
 \makecell[c]{$\left(\ydiagram{2}, \mydot \right)$ \\[1ex] $\left(\ydiagram{3}, \mydot \right)$\\[1ex] $\left(\ydiagram{5,2}, \mydot \right)$} 
 &\makecell[c]{$\left(\ydiagram{2,1}, \mydot \right)$ \\[1ex] $\left(\ydiagram{4,1}, \mydot \right)$
 }
 & \makecell[c]{ $\left(\ydiagram{3,1}, \mydot \right)$ \\[1ex] $\left(\ydiagram{4,2}, \mydot \right)$}
 \\
     \hline
        $q = 1,7 \mod 8$ & 1  &$1+\sqrt{2}$ & $2+\sqrt{2}$&$2(1+\sqrt{2})$& $3 + 2 \sqrt{2}$\\
        $q = 3,5 \mod 8$ & 1  &$1-\sqrt{2}$ & $2-\sqrt{2}$&$2(1-\sqrt{2})$& $3 - 2 \sqrt{2}$              \\
        \hline
    \end{tabular}
    \caption{Quantum dimension of the preserved subcategory ${\rm Rep}[SU(3)_5]$ in the minimal models $\cW_3(8,q)$. For reasons of space we have omitted writing the complex conjugate representations as $d_\cL = d_{\overline{\cL}}$ in a unital fusion category. } 
    \label{tab:W3p8}
\end{table}

\begin{align*}
\includegraphics[width=\linewidth]{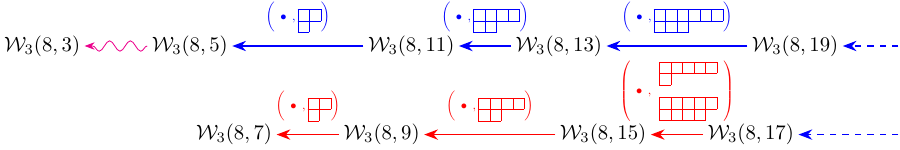}
\end{align*}
In the cases $\cW_3(7,4)$ and $\cW_3(8,5)$ the flows may be extended to $q=3$. Yet, we have not found a possible relevant primary that would trigger the deformation.

\subsection*{\boldmath \texorpdfstring{Appendix C: Detailed analysis  for $\mathcal{W}_4$}{Appendix C: Detailed analysis  for W4}}

\subsubsection{\textbf{General proofs}}

\paragraph{\textbf{Commutation condition.}}
Consider a generic right field  $\phi\in SU(4)_{q-4}$ parametrized by $(k_1-k_2,k_2-k_3,k_3)$ (i.e.\ by Young diagram of row lengths $[k_1,k_2,k_3]$)
 and a line $\cL_\sigma \in \Rep[SU(3)_{p-3}]$ parametrized as by $\sigma = (l_1-l_2,l_2-l_3,l_3)$ (i.e.\ by row lengths $[l_1,l_2,l_3]$). 
Computing explicitly the commutation condition \eqref{simplecomm} in the minimal model $\cW_4(p,q)$ one finds: (with $t = q/p$) 
\be \begin{split}
d_{\cL_\sigma} - \frac{S_{\sigma \phi}}{S_{0 \phi}} = -\frac{(1-i) \csc ^6(\pi  t) \sec ^2(\pi  t)}{4 \sin (2 \pi  t)+2} \sin (\pi  (l_1+3) t) \sin (\pi  (l_1-l_2+1) t) \sin (\pi  (l_2+2) t)\\ \times \sin (\pi  (l_1-l_3+2) t) \sin (\pi  (l_2-l_3+1) t) \sin (\pi  (l_3+1) t) \\
\times\color{blue}\sin^2(\frac{1}{4} \pi  (k_1 (-3 l_1+l_2+l_3)+k_2 (l_1-3 l_2+l_3)+k_3 (l_1+l_2-3 l_3)))\end{split}\ee
Analogously to the $\cW_3$ case, the terms in black never vanish due to the ranges $l_3\leq l_2\leq l_1 \leq p-4$. Instead, for the blue term, one directly computes
$$ k_1 (-3 l_1+l_2+l_3)+k_2 (l_1-3 l_2+l_3)+k_3 (l_1+l_2-3 l_3) \mod 4 =  (k_1+k_2+k_3)(l_1 + l_2 + l_3)\mod 4.$$
Then:
\begin{itemize}
    \item for $\sum k_i \in 4\bZ$, that is, for $\phi \in PSU(4)_{q-4}$, the full $\Rep[SU(4)_{p-4}]$ is preserved;
    \item  for $\sum k_i \in 2\bZ$, that is, for $\phi \in (SU(4)/\bZ_2)_{q-4}$, only $\Rep[(SU(4)/\bZ_2)_{p-4}]$, being the subcategory of even $4$-ality with $\sum l_i \in 2\bZ$ is preserved;
    \item for generic $\phi$, only $PSU(4)_{p-4}$ survives;
\end{itemize}
as claimed in the main text. 
\paragraph{\textbf{Matching of quantum dimensions.}}
The quantum dimension of a line in $\Rep[SU(4)_{p-4}]$ is
\be\begin{split} d_{\cL} =\frac{(-1)^{l_1+l_2+l_3} \csc ^6(\pi  t) \sec ^2(\pi  t)}{8 \cos (2 \pi  t)+4} \sin (\pi  (l_1+3) t) \sin (\pi  (l_1-l_2+1) t) \sin (\pi  (l_2+2) t) \\ \times\sin (\pi  (l_1-l_3+2) t) \sin (\pi  (l_2-l_3+1) t) \sin (\pi  (l_3+1) t).\end{split}\ee
Specifying $q = k p \pm i$ with $k\in \bZ$ we find
\be\begin{split} d_{\cL}[\cW_3(p,kp \pm i)] &= \tfrac{(-1)^{(3k+1)(l_1 + l_2 + l_3)+2k } }{4( 2\cos (\tfrac{ \pm 2\pi  i}{p})+1) \cos^2(\tfrac{\pm \pi  i}{p}) \sin ^6(\tfrac{\pm \pi  i}{p}) } \sin (\tfrac{ \pm \pi  i(l_1+3)}{p}) \sin (\tfrac{\pm \pi  i(l_1-l_2+1) }{p}) \sin (\tfrac{ \pm \pi  i(l_2+2)}{p})\\ &\qquad \qquad \times \sin (\tfrac{ \pm \pi  i(l_1-l_3+2)}{p}) \sin (\tfrac{\pm \pi  i(l_2-l_3+1) }{p}) \sin (\tfrac{\pm \pi  i(l_3+1) }{p})  \\
&= \tfrac{(-1)^{l_1 + l_2 + l_3 }  {\color{blue} (-1)^{k(l_1 + l_2 + l_3)  }}  }{4( 2\cos (\tfrac{  2\pi  i}{p})+1) \cos^2(\tfrac{ \pi  i}{p}) \sin ^6(\tfrac{ \pi  i}{p}) } \sin (\tfrac{  \pi  i(l_1+3)}{p}) \sin (\tfrac{ \pi  i(l_1-l_2+1) }{p}) \sin (\tfrac{ \pi  i(l_2+2)}{p})\\ &\qquad \qquad \times \sin (\tfrac{  \pi  i(l_1-l_3+2)}{p}) \sin (\tfrac{ \pi  i(l_2-l_3+1) }{p}) \sin (\tfrac{ \pi  i(l_3+1) }{p}).
\end{split}\ee
Therefore, if $k\in \bZ$ and hence also $i\in \bZ$, the quantum dimension match across  $\cW_3(p,kp \pm i)$, but due to the extra sign, $(-1)^{k(l_1 + l_2 + l_3)}$ are do no take the same values over conjugacy classes of $p$ -- rather they are invariant only $\mod 2p$. It also follows directly that $d_{\cL}$ with $\cL\in SU(4)/\bZ_2$ is matched between the flows with half integer $k$. Along those, the odd $N$-ality lines differ exactly by a minus sign, as claimed in the main text.

\subsection{\boldmath \texorpdfstring{\textbf{Detailed illustration for }$p \leq 8$}{{Detailed illustration for} p < 8}}

\paragraph{\boldmath $\cW_4(4,q)$}
In this case there is no category of preserved lines, and therefore no anomalies can be matched. It is analogous to $\cW_3(3,q)$ and $\cW_3(3,q)$.

\paragraph{\boldmath $\cW_4(5,q)$}
For $p =5$ the preserved subcategory is $\Rep(SU(4)_1) \equiv \bZ_4$. The only non-trivial quantum dimensions are:
$$d_{{\scalebox{0.6}{\ydiagram{1}}}}[\cW_{5, q }] = (-1)^q\comma \qquad d_{{\scalebox{0.6}{\ydiagram{1,1}}}}[\cW_{5, q }] =1 $$
as in Table \ref{tab:W4p5}.
\begin{table}[]
    \centering
    \begin{tabular}
    {|c||c|c|}
     \hline
 \multicolumn{3}{|c|}{q-dimensions  in $\cW_4(5,q)$ of  ${\rm Rep}[SU(4)_1]\equiv \bZ_4$} \\
\hline   & $\left(\mydot, \mydot \right) \qquad  \left(\ydiagram{1,1}, \mydot \right)$ 
 & $\left(\ydiagram{1}, \mydot \right) \qquad \left(\ydiagram{1,1,1}, \mydot \right)$  \\[0.2cm]
     \hline
        $q = 1,3,7,9 \mod 10 $ & 1  &$-1$\\
       $q = 2,4,6,8 \mod 10 $ & 1  &$1$ \\  
        \hline
    \end{tabular}
    \caption{Quantum dimension of the preserved $\bZ_4$  symmetry in the minimal models $\cW_4(5,q)$. } 
    \label{tab:W4p5}
\end{table}
Hence the allowed flows are:
$$
\includegraphics[width = \linewidth]{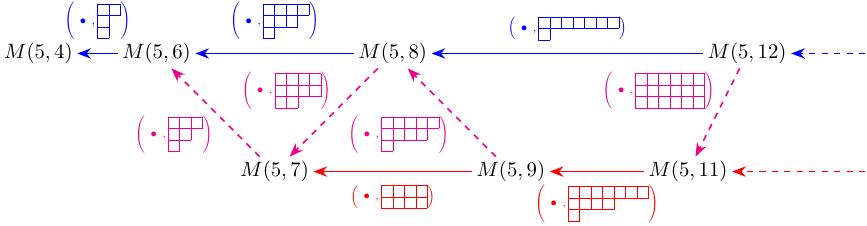}
$$
Note that the case with only $\bZ_4$ symmetry is special as discussed in the main text. The dashed flows are the one preserving only the $\bZ_2\subset \bZ_4$ generated by  $\Big({\scalebox{0.6}{\ydiagram{1,1}}}, \mydot \Big)$.

\paragraph{\boldmath $\cW_4(6, q)$} \begin{table}[]
    \centering
    \begin{tabular}
    {|c||c|c|c|}
     \hline
 \multicolumn{4}{|c|}{q-dimensions  in $\cW_4(6,q)$ of  ${\rm Rep}[SU(4)_2]$} \\
\hline   & \makecell[c]{$\left(\mydot, \mydot \right)$ \\[1ex] $\left(\ydiagram{2,2,2}, \mydot \right)$} 
\makecell[c]{$\left(\ydiagram{2}, \mydot \right)$\\[1ex] $\left(\ydiagram{2,2}, \mydot \right)$}
 & \makecell[c]{$\left(\ydiagram{1}, \mydot \right)$ \\[1ex] $\left(\ydiagram{1,1,1}, \mydot \right)$} \makecell[c]{$\left(\ydiagram{2,1}, \mydot \right)$ \\[1ex] $\left(\ydiagram{2,2,1}, \mydot \right)$} &\makecell[c]{$\left(\ydiagram{1,1
 }, \mydot \right)$} \makecell[c]{$\left(\ydiagram{2,1,1}, \mydot \right)$}  \\[0.2cm]
     \hline
        $q = 1,11 \mod 12 $ & 1  &$-\sqrt{3}$ & $2$\\
       $q = 5,7 \mod 12 $ & 1  &$\sqrt{3}$ & $2$\\         
        \hline
    \end{tabular}
    \caption{Quantum dimension of the preserved subcategory ${\rm Rep}[SU(3)_4]$ in the minimal models $\cW_4(6,q)$. } 
    \label{tab:W4p6}
\end{table}
In the case of $q=6$ the quantum dimension are reported in Table \ref{tab:W4p6}. The adjoint deformations by $PSU(4)_{q-3}$ preserve $SU(4)_{2}$, while the ones by generic primaries $SU(4)_{q-3}$ only preserve the algebra generated by the even lines $\Big\{\Big(\scalebox{0.6}{\ydiagram{1,1}},\mydot\Big), \Big(\scalebox{0.6}{\ydiagram{2,1,1}},\mydot\Big) , \bZ_4 \Big\}.$
The resulting flows are:
$$
\includegraphics[width = \linewidth]{W4p6.pdf}
$$

\paragraph{\boldmath$\cW_4(7,q)$}
The quantum dimension are listed in \ref{tab:W4p7}. The adjoint deformation in $PSU(4)_{q-4}$ commutes with the preserved subcategory  $SU(4)_{3}$, while the one in $SU(4)_{q-4}$ commutes with the smaller $PSU(3)_{3}$. 
\begin{table}[]
    \centering
    \begin{tabular}
    {|c||c|c|c|c|c|c|c|}
     \hline
 \multicolumn{8}{|c|}{q-dimensions  in $\cW_4(7,q)$ of  ${\rm Rep}[SU(4)_3]$} \\
\hline   & \makecell[c]{$(\mydot,\mydot)$\\ $\left(\ydiagram{3,3}, \mydot \right)$} &   
$\left(\ydiagram{3}, \mydot \right)$ & \makecell[c]{$\left(\ydiagram{1}, \mydot \right)$\\ $\left(\ydiagram{3,2}, \mydot \right)$}& \makecell[c]{$\left(\ydiagram{2}, \mydot \right)$\\ $\left(\ydiagram{3,1}, \mydot \right)$}&\makecell[c]{$\left(\ydiagram{1,1}, \mydot \right)$\\ $\left(\ydiagram{2,2}, \mydot \right)$} &{$\left(\ydiagram{2,1}, \mydot \right)$} &{$\left(\ydiagram{2,1,1}, \mydot \right)$} \\
\hline 
$q=1, 13 \mod 14$& $1$ & $-1$&$-r_1$& $r_1$&$s_1$& $-t_1$&$t_1$ \\
$q=2, 12 \mod 14$& $1$ & $1$&$r_2$& $r_2$&$s_2$&$t_2$&$t_2$\\ 
$q=3, 11 \mod 14$& $1$ & $-1$&$-r_3$& $r_3$&$s_3$&$-t_3$&$t_3$\\
$q=4, 10 \mod 14$& $1$ & $1$&$r_3$& $r_3$&$s_3$&$t_3$&$t_3$\\
$q=5, 9 \mod 14$& $1$ & $-1$&$-r_2$& $r_2$&$s_2$&$-t_2$&$t_2$\\
$q=6, 8 \mod 14$& $1$ & $1$&$r_1$& $r_1$&$s_1$&$t_1$&$t_1$\\
\hline
    \end{tabular}
    \caption{Quantum dimension of the preserved subcategory ${\rm Rep}[SU(4)_3]$ in the minimal models $\cW_4(7,q)$. The values of $r_i,s_i,t_i$ are reported in \eqref{thirdroots}.} 
    \label{tab:W4p7}
\end{table}
The resulting flows are:
\begin{align*}
    \includegraphics[width = \linewidth]{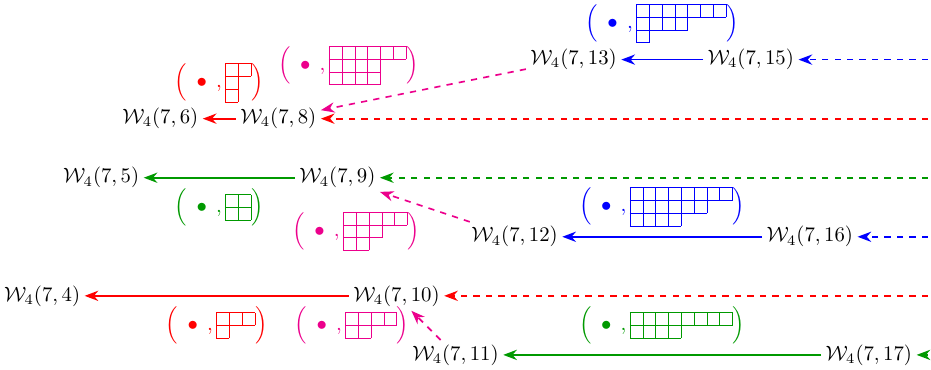}
\end{align*}
It is clear that no other flows preserving the even smaller $PSU(4)$ category are allowed. The values $r_i,s_i,t_i$ are the roots of third order polynomials \eqref{thirdroots}.

\paragraph{\boldmath $\cW_4(8,q)$}
\begin{table}[]
    \centering
   \hbox{\scalebox{.95}{\begin{tabular}
    {|c||c|c|c|c|c|c|c|c|}
     \hline
 \multicolumn{8}{|c|}{q-dimensions  in $\cW_4(8,q)$ of  ${\rm Rep}[SU(4)_4]$} \\
\hline   & \makecell[c]{$(\mydot,\mydot)$\\ \\ $\left(\ydiagram{4}, \mydot \right)$\\ \\$\left(\ydiagram{4,4}, \mydot \right)$} &   \makecell[c]{$\left(\ydiagram{1}, \mydot \right)$\\\\ $\left(\ydiagram{3}, \mydot \right)$ \\ \\$\left(\ydiagram{4,1}, \mydot \right)$\\\\ $\left(\ydiagram{4,3}, \mydot \right)$}& \makecell[c]{$\left(\ydiagram{2}, \mydot \right)$\\ $\left(\ydiagram{1,1}, \mydot \right)$\\$\left(\ydiagram{4,2}, \mydot \right)$\\$\left(\ydiagram{3,3}, \mydot \right)$\\ $\left(\ydiagram{4,1,1}, \mydot \right)$}&\makecell[c]{$\left(\ydiagram{2,1}, \mydot \right)$\\ $\left(\ydiagram{3,1,1}, \mydot \right)$\\ $\left(\ydiagram{4,2,1}, \mydot \right)$\\ $\left(\ydiagram{3,2}, \mydot \right)$} &\makecell[c]{$\left(\ydiagram{2,1,1}, \mydot \right)$\\ \\$\left(\ydiagram{3,1}, \mydot \right)$ \\ \\ $\left(\ydiagram{4,3,1}, \mydot \right)$} & \makecell[c]{$\left(\ydiagram{2,2}, \mydot \right)$\\$\left(\ydiagram{4,2,2}, \mydot \right)$} & $\left(\ydiagram{3,2,1}, \mydot \right)$\\
\hline 
$q=1, 15$& $1$ & $-\sqrt{2(2+ \sqrt{2})}$&$2+ \sqrt{2}$& $-\sqrt{20+14\sqrt{2}}$&$3+2 \sqrt{2}$& $2 (1+\sqrt{2})$&$4 (1+\sqrt{2})$ \\
$q=3, 13$& $1$ & $\sqrt{2(2- \sqrt{2})}$&$2- \sqrt{2}$& $-\sqrt{20-14\sqrt{2}}$&$3-2 \sqrt{2}$& $2 (1-\sqrt{2})$&$4 (2-\sqrt{2})$ \\ 
$q=5, 11$& $1$ & $-\sqrt{2(2- \sqrt{2})}$&$2-\sqrt{2}$& $\sqrt{20-14\sqrt{2}}$&$3-2 \sqrt{2}$& $2 (1-\sqrt{2})$&$4 (2-\sqrt{2})$ \\
$q=7, 9 $& $1$ & $\sqrt{2(2+ \sqrt{2})}$&$2+\sqrt{2}$& $\sqrt{20+14\sqrt{2}}$&$3+2 \sqrt{2}$& $2 (1+\sqrt{2})$&$4 (1+\sqrt{2})$\\
\hline
    \end{tabular}}}
    \caption{Quantum dimension of the preserved subcategory ${\rm Rep}[SU(4)_4]$ in the minimal models $\cW_4(8,q)$. The values of $q$ are intended $\mod 16$.} 
    \label{tab:W4p8}
\end{table}
In this case the preserved subcategory is $SU(4)_4$. The quantum dimensions are reported in Table \ref{tab:W4p8}. The allowed flows are: 
\begin{align*}
    \includegraphics[width = \linewidth]{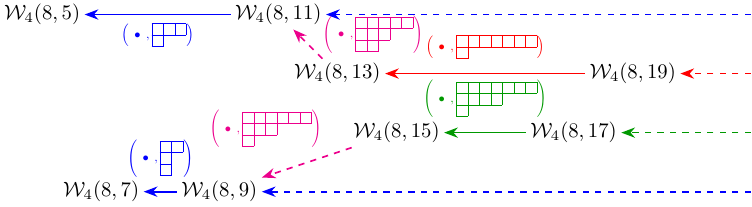}
\end{align*}

\subsection{Appendix D: Comparison with known RG flows.}
In this Appendix we discuss (some of the) known integrable deformations and flows between minimal models and compare them to our findings.

\paragraph{\textbf{Virasoro.}}
In the Virasoro case, we have the three famous integrable deformations $\phi_{[1,2]},\phi_{[1,3]}, \phi_{[1,5]}$ corresponding to $k=1/2, k=1,k=2$\cite{Zamolodchikov:1987ti, Dorey:2000zb}. In \cite{Ambrosino:2025myh,Ambrosino:2025pjj,Ambrosino:2025yug} evidence of integrability for $\phi_{[1,2k+1]}$ has been provided in the form of infinitely many new (non-local) conserved charges, at least for specific values of $p$ and $q$.

\paragraph{\boldmath $\cW_3$ \textbf{algebra.}}
There are four known integrable deformations of $\cW_3$ minimal models \cite{Vaisburd:1994hg}. These are: $$F_{(22|11)}\leftrightarrow \Big(\mydot, \scalebox{0.7}{\ydiagram{2,1}}\Big)\comma \qquad F_{(21|11)}\leftrightarrow \Big(\mydot, \scalebox{0.7}{\ydiagram{1}}\Big) \comma \qquad  F_{(11|21)}\leftrightarrow \Big( \scalebox{0.7}{\ydiagram{1}}, \mydot\Big) 
\comma \qquad F_{(41|11)}\leftrightarrow \Big(\mydot, \scalebox{0.7}{\ydiagram{3}}\Big)$$.
The first one of those is the adjoint deformation coming from reducing $A_2^{(1)}$. Indeed, it is known \cite{Lukyanov:1990tf,Poghosyan:2022mfw,Poghosyan:2022ecv}that the adjoint deformation induces the RG flows  between the unitary minimal models of the $W_3(p,p+1)\xrightarrow{\Big(\mydot, \scalebox{0.7}{\ydiagram{2,1}}\Big)} W_3(p,p-1)$, corresponding to \eqref{W3flows}  with $k =i=1$ . The pair $F_{(11|21)},F_{(21|11)}$ is the analogous of the pair $(1,2), (2,1)$ of Virasoro, and in our convention of always having the preserved subcategory to be the left one it corresponds to the $PSU(3)$ preserving flow via the fundamental weight $\omega_1$. $F_{(41|11)}$ corresponds to the reduction from $D_4^{(3)}\simeq G_2^{(1)}$. Indeed, in agreement to what we also find, we are not aware of any integrable massless flows purely along these other deformations.

\paragraph{\boldmath $\cW_4$ \textbf{algebra.}} There are 3 known deformations of $\cW_4$ models\cite{Vaisburd:1994hg}:
$$ 
F_{(212|111)} \leftrightarrow \bigg(\mydot, \scalebox{0.7}{\ydiagram{2,1,1}}\bigg)\comma \qquad F_{(121|111)} \leftrightarrow \Big(\mydot, \scalebox{0.7}{\ydiagram{1}}\Big)\comma \qquad F_{(131|111)} \leftrightarrow \bigg(\mydot, \scalebox{0.7}{\ydiagram{1,1}}\bigg)
$$
corresponding respectively to reduction from $A_3^{(1)}$, $B_3^{(1)}$ and $A_5^{(2)}$. They correspond to the simplest representative from our three special classes of deformations preserving $SU(4)_{p-4}$, $PSU(4)_{p-4}$ and $PSU(4)/\bZ_2$ respectively.

\paragraph{\boldmath \textbf{General case.}} In the general case, the following universal integrable flows are known\cite{Lukyanov:1990tf}: 
$$ 
\frac{SU(N)_\kappa \times SU(N)_{\rho}}{SU(N)_{\kappa+\rho}} \xrightarrow[]{\phi^{Adj}_{1,1}} \frac{SU(N)_{\kappa-\rho} \times SU(N)_{\rho}}{SU(N)_{\kappa}}.
$$
For $\rho = 1$ these correspond to our flows with $k=1$ and generic $N$, and the coset field $\phi^{\rm Adj}_{1,1}$ is exactly $\phi = (\mydot, \rm Adj)$ in our conventions. The ones with $\rho=2$ have been studied in \cite{Gaberdiel:2026sfg}. 
\paragraph{\textbf{An interesting conjecture.}}
It would be intriguing to propose a conjecture based on extrapolation of what we have seen in this paper. Namely, one may try to prove the deformation by the simplest representative from each of the classes corresponding to $(SU(N)/\bZ_K)_{p-N}$ with $\bZ_K$ proper subgroup of $\bZ_N$, is integrable.

\end{document}

%% file: preambleLetter.tex
\pdfoutput=1
\usepackage{mathtools, amsfonts, amsthm, latexsym, amssymb,dsfont}
\usepackage[T1]{fontenc}
\usepackage{microtype}

\usepackage{times}
\usepackage{slashed}
\usepackage{hyperref}
\hypersetup{pdftex,
    pdfauthor={F. Ambrosino and T. Prochazka},
    colorlinks=true, citecolor=blue, urlcolor=blue,
    pdfstartview={FitH}, 
pdftitle={WNflows}, 
linkcolor=blue, 
citecolor=blue, 
filecolor=magenta, 
}

\usepackage{lipsum}
\usepackage{amsmath}
\usepackage{tikz}
\usepackage{subcaption}
\usepackage{bbm}

\usepackage{color}
\usepackage{physics}

\newcommand{\cL}{{\mathcal L}}

\newcommand{\cI}{{\mathcal I}}
\newcommand{\cH}{{\mathcal H}}

\newcommand{\bZ}{{\mathbb Z}}

\def\be{\begin{equation}}
\def\ee{\end{equation}}
\def\comma{\,,}
\def\period{\,.}

\def\({\left(}
\def\){\right)}

\usepackage{braket}
\usepackage{subcaption}

\usepackage{ytableau} 
\ytableausetup{boxsize=1em, centertableaux} 

\makeatletter 
    
\renewcommand\onecolumngrid{
\do@columngrid{one}{\@ne}%
\def\set@footnotewidth{\onecolumngrid}
\def\footnoterule{\kern-6pt\hrule width 1.5in\kern6pt}%
}

\renewcommand\twocolumngrid{
        \def\footnoterule{
        \dimen@\skip\footins\divide\dimen@\thr@@
        \kern-\dimen@\hrule width.5in\kern\dimen@}
        \do@columngrid{mlt}{\tw@}
}%

\makeatother

\usepackage{commands}

\usepackage{makecell}